\documentclass[aps,pre,twocolumn,superscriptaddress,nofootinbib]{revtex4-1}
\usepackage{graphicx}
\usepackage{amsmath}

\begin{document}

\title{Interspecies evolutionary dynamics mediated by public goods\\
in bacterial quorum sensing}

\author{Eduardo J. Aguilar}
\affiliation{Instituto de Ci\^encia e Tecnologia,
Universidade Federal de Alfenas,
Rod.\ Jos\'e Aur\'elio Vilela, 11999, 37715-400 Po\c cos de Caldas - MG,
Brazil}

\author{Valmir C. Barbosa}
\email[]{valmir@cos.ufrj.br}
\affiliation{Programa de Engenharia de Sistemas e Computa\c c\~ao, COPPE,
Universidade Federal do Rio de Janeiro,
Centro de Tecnologia, Sala H-319, 21941-914 Rio de Janeiro - RJ, Brazil}

\author{Raul Donangelo}
\affiliation{Instituto de F\'\i sica, Facultad de Ingenier\'\i a,
Universidad de la Rep\'ublica,
Julio Herrera y Reissig 565, 11.300 Montevideo, Uruguay}
\affiliation{Instituto de F\'\i sica,
Universidade Federal do Rio de Janeiro,
Centro de Tecnologia, Bloco A, 21941-909 Rio de Janeiro - RJ, Brazil}

\author{Sergio R. Souza}
\affiliation{Instituto de F\'\i sica,
Universidade Federal do Rio de Janeiro,
Centro de Tecnologia, Bloco A, 21941-909 Rio de Janeiro - RJ, Brazil}
\affiliation{Departamento de F\'\i sica, ICEx,
Universidade Federal de Minas Gerais,
Av.\ Ant\^onio Carlos, 6627, 31270-901 Belo Horizonte - MG, Brazil}

\begin{abstract}
Bacterial quorum sensing is the communication that takes place between bacteria
as they secrete certain molecules into the intercellular medium that later get
absorbed by the secreting cells themselves and by others. Depending on cell
density, this uptake has the potential to alter gene expression and thereby
affect global properties of the community. We consider the case of multiple
bacterial species coexisting, referring to each one of them as a genotype and
adopting the usual denomination of the molecules they collectively secrete as
public goods. A crucial problem in this setting is characterizing the
coevolution of genotypes as some of them secrete public goods (and pay the
associated metabolic costs) while others do not but may nevertheless benefit
from the available public goods. We introduce a network model to describe
genotype interaction and evolution when genotype fitness depends on the
production and uptake of public goods. The model comprises a random graph to
summarize the possible evolutionary pathways the genotypes may take as they
interact genetically with one another, and a system of coupled differential
equations to characterize the behavior of genotype abundance in time. We study
some simple variations of the model analytically and more complex variations
computationally. Our results point to a simple trade-off affecting the long-term
survival of those genotypes that do produce public goods. This trade-off
involves, on the producer side, the impact of producing and that of absorbing
the public good. On the non-producer side, it involves the impact of absorbing
the public good as well, now compounded by the molecular compatibility between
the producer and the non-producer. Depending on how these factors turn out,
producers may or may not survive.
\end{abstract}

\maketitle

\section{Introduction}
\label{sec:int}

Bacterial quorum sensing (QS) is the general denomination for a variety of
mechanisms whereby bacteria communicate with one another by the secretion and
uptake of molecules that get diffused in the intercellular medium. The
absorption of such molecules by the cells is important for the regulation of
some genes and thus affects survival and successful proliferation, as well as
many functional traits, depending on cell density. These include a bacterial
species' propensity to establish symbioses or to aggregate into biofilms, and
also its virulence and motility. Although bacterial QS has been around as a
research topic for about five decades, only much more recently has sufficient
evidence accumulated. Interest in it is now widespread and a number of useful
reviews are available \cite{mb01,hjn16,pb16,pgg16,sspt17,wdg17}. Whereas
initial interest was sparked by the curious phenomenon of bioluminescence
arising in certain marine bacteria \cite{nph70,ghu79}, the current focus is on
the role of QS in the immunological effects of the mammalian gut microbiome
\cite{gc18,hz18} and in combating bacterial virulence \cite{rb12,d18}.

When only one single species is involved in bacterial QS, a cell's uptake,
though comprising mainly the molecules that the cell itself secretes, may also
include those of the other cells nearly indistinguishably. Such molecules are
called autoinducers, in allusion to the fact that they act as gene regulators
for the species. In settings allowing for multiple species, it is in principle
conceivable that the autoinducers secreted by members of a species be absorbable
also by those of another. Depending on the species involved, the  autoinducers
they exchange can be expected to affect cells of different species differently,
perhaps acting as proper autoinducers in some cases and simply as sources of
nutrients or even toxicity in others. A cell's secretions, moreover, may include
molecules other than the ones directly involved in autoinduction. Commonly, all
secreted molecules are collectively referred to as public goods (cf., e.g.,
\cite{wdg17}). In evolutionary terms, and depending on the metabolic costs
involved, the production and uptake of public goods can affect a cell's fitness
significantly.

Despite the much earlier developments in the field, bacterial QS seems to have
been the object of mathematical modeling for no longer than about two decades.
Many of the modeling efforts have counted on the close participation of
researchers involved with experimentation, resulting in detailed analytical
accounts of specific QS processes \cite{pgg16}, including the combinatorial
response to intercellular signaling \cite{cpmgsidb14}. Others have taken a
somewhat more distanced stance and addressed, e.g., transitions in bacterial QS
\cite{yb16}, and biofilm \cite{vaga18} and pattern \cite{rhss16,bf18} formation.
Still others have taken up the basic tenets of QS and sought to characterize
higher-level phenomena occurring not only in bacteria but also in other systems
in which QS can be said to play a central role. Such phenomena include
synchronization in networks \cite{rs10}, oscillations in coupled systems
\cite{vcs19}, the appearance of dense aggregates of active particles
\cite{fss20}, and phase separation in colloidal mixtures \cite{sk20}. To the
best of our knowledge, each of these models has considered one single bacterial
species, taking diversity into account only insofar as it stems from a few of
that species' strains.

Here we introduce a model of bacterial QS that focuses on how multiple species
coexist and evolve in the presence of the public goods they produce. In order to
avoid the many complications involved in bacterial taxonomy \cite{ra15}, we
henceforth refer to genotypes as the basic units of diversity. Given two
genotypes, one of them might be considered merely a strain of the other or they
might both refer to totally distinct species. In our model, all possible
genotypes of a given length are considered concomitantly. The complete QS system
is represented by a random graph whose nodes stand for genotypes and whose edges
exist with probabilities that aim to reflect the mutations that occur as cells
undergo binary fission (the most common mechanism of bacterial proliferation,
though several others exist \cite{a05,b16}) and horizontal gene transfer (HGT,
the transfer of genetic material between individuals \cite{hbh17}). These
probabilities are parameterized so that sampling from the random graph may
result in a graph with more or fewer edges and therefore more or less
far-reaching effects of localized random variation. 

In addition to the structural component given by this random graph, our model
includes a system of coupled differential equations describing how the
genotypes' abundances vary with time. These equations depend on the same
parameterization as the random graph. Crucially, they also depend on which
genotypes are producers of public goods, as well as on whether and how
absorption affects each individual genotype. All these aspects of public-good
production and uptake contribute to each genotype's fitness in the QS system and
can be quantified in such a way that a great variety of global configurations
can be studied.

This model continues earlier work \cite{bds12} in which we added network
structure to the so-called quasispecies models of the dynamics of prebiotic
molecules and RNA viruses \cite{e71,es77,be06,d09,la10,mlcgm10}, and also work
in which we modeled the network dynamics of autoimmunity and immunodeficiency  
\cite{bds15,bds16} and of the evolution of eukaryotic cellular division
\cite{bds18}. We now proceed by giving the model's details in
Sec.~\ref{sec:mod} and analyzing a special case thereof in Sec.~\ref{sec:sc1}.
Results are given in Sec.~\ref{sec:res} and discussed in Sec.~\ref{sec:dis}. We
conclude in Sec.~\ref{sec:con} and give directions for further analysis in the
Appendix.

\section{Model}
\label{sec:mod}

We represent each bacterial gene by a sequence of $B$ binary digits ($0$'s or
$1$'s), each standing for a nucleotide. A bacterial genotype is represented by a
sequence of $L$ binary digits, provided $L\ge B$ and that $L$ is a multiple of
$B$. A genotype is then viewed as a sequence of $L/B$ genes. Such
representations are of course oversimplistic in more than one sense, but even so
they provide the necessary means to model the two main sources of variation in
bacterial genotypes, viz., mutations during binary fission (at the level of the
nucleotide) and HGT (at the level of the gene). Henceforth we use $N$ to denote
the set of all $2^L$ genotypes. Genotypes are numbered $0$ through $2^L-1$.

Our focus is on the effects of public-good production and consumption by the
genotypes, which we model using parameters $\mu_j$ and $\sigma_{ji}$, as
follows. For each genotype $j$, $\mu_j$ takes its value from $\{0,1\}$ and
indicates whether $j$ is a producer of public goods ($\mu_j=1$) or not
($\mu_j=0$). For each genotype pair $i,j$, $\sigma_{ji}$ takes its value from
the interval $[-1,1]$ and is used to indicate the degree to which consuming
public goods produced by genotype $j$ can impact the rate at which genotype $i$
is capable of multiplying. This degree ranges from a full contribution toward a
slower pace ($\sigma_{ji}=-1$) to a full contribution toward a faster pace
($\sigma_{ji}=1$), with neutrality in between ($\sigma_{ji}=0$). For $j=i$, the
biological process in question includes that of bacterial autoinduction.

Whenever $\mu_j\sigma_{ji}\neq 0$, we model the consumption by $i$ of the public
goods produced by $j$ as occurring with probability $u_{ji}=u^{J_{ij}}$, where
$u<1$ is a base probability of the model and $J_{ij}$ is the number of
homologous genes at which genotypes $i$ and $j$ differ. Expressing $u_{ji}$ in
this way takes into account the need for molecular compatibility between
organisms of genotype $j$ and those of genotype $i$, and the fact that such
compatibility occurs at the level of the genes of $j$ and $i$. Thus, lower
values of $J_{ji}$ indicate higher compatibility. $J_{ji}=0$, in particular,
holds when $j=i$, so autoinduction occurs with probability $1$ whenever
$\mu_j=1$.

As will become apparent when we give the model's dynamical equations in
Sec.~\ref{sec:dyn}, we make another strong simplifying assumption, now regarding
the production and consumption of public goods by the genotypes. The assumption
is that setting $\mu_j=1$ for some genotype $j$ makes the public goods it
produces available to all genotypes in sufficient concentration at all times.
That is to say, we make no provision to model the dynamics of public-good
production, diffusion, or consumption. Our equations are concerned solely with
the dynamics of genotype proliferation, given a static backdrop of overall
public-good availability.

\subsection{Fitness of a genotype}
\label{sec:fit}

The quantification of an individual's fitness in studies of evolutionary
dynamics can be challenging. Ideally, it should rely on the identification of
traits exerting a quantifiable influence on gene spread through the population.
This is rarely possible, so it is common for the difficulty to be sidestepped
with the adoption of a proxy, as some form of distance to an agreed-upon wild
type or the experimental counting of individuals as proliferation unfolds.

However, the case at hand is exceptional in that, clearly, a genotype can be
said to be as fit as its total consumption of public goods is beneficial to it.
That is, if genotype $j$ is a producer of public goods (i.e., $\mu_j=1$), then
its contribution to the fitness of genotype $i$ should be an increasing function
of $\sigma_{ji}u^{J_{ij}}$. It should also depend on how abundant genotype $j$
is in the entire population, which immediately makes the fitness of $i$
time-dependent. Letting $X_j$ denote the absolute abundance of genotype $j$ at a
certain instant and
\begin{equation}
x_j=\frac{X_j}{\sum_{k\in N}X_k}
\label{xj}
\end{equation}
its relative abundance (so $\sum_{j\in N}x_j=1$), we express the instantaneous
fitness of $i$ by an increasing function of the convex combination of
$\mu_j\sigma_{ji}u^{J_{ij}}$ over $N$ that uses $x_j$ as the weight of genotype
$j$. We denote this combination by $d_i$, so
\begin{equation}
d_i=\sum_{j\in N}\mu_j\sigma_{ji}u^{J_{ij}}x_j,
\label{di}
\end{equation}
and denote the fitness of genotype $i$ by $f_i$, defining it as $f_i=2^{d_i}$.
Clearly, $-1\le d_i\le 1$, whence $0.5\le f_i\le 2$. For $d_i=0$ (the combined
public goods instantaneously available to $i$ are neither beneficial nor
detrimental to its ability to multiply), we have $f_i=1$. This is convenient in
view of how we incorporate fitness values into the model's dynamical equations
(cf.\ Sec.~\ref{sec:dyn}).

\subsection{Network structure}
\label{sec:str}

A further important element in our model is a substrate on which genotypes can
transform into one another via the mechanisms afforded by mutations and HGT. The
substrate we use is a random graph of node set $N$ (the set of all $2^L$
distinct genotypes on $L$ nucleotides) and undirected edges. The existence of
an edge between any two genotypes $i$ and $j$ depends on two base probabilities,
$p$ and $r$, and has probability
\begin{equation}
\pi_{ij}=p^{H_{ij}}+r^{J_{ij}}-p^{H_{ij}}r^{J_{ij}}.
\label{pij}
\end{equation}
In this expression, $H_{ij}$ is the Hamming distance between genotypes $i$ and
$j$, i.e., the number of nucleotides at which they differ. $J_{ij}$, already
introduced at the beginning of Sec.~\ref{sec:mod}, is the number of homologous
$B$-nucleotide genes at which $i$ and $j$ differ.

Given the value of $p$, increasing $H_{ij}$ makes it less likely for the edge
joining $i$ and $j$ to exist, reflecting the fact that the transformation of $i$
into $j$ (or conversely) by co-occurring mutations at $H_{ij}$ nucleotides
becomes less likely as well. Similarly, given the value of $r$, increasing
$J_{ij}$ makes joining $i$ and $j$ by an edge less likely, now reflecting the
also less likely transformation of $i$ into $j$ (or conversely) by the
co-occurring horizontal transfer of $J_{ij}$ genes. The expression in
Eq.~(\ref{pij}) comes from assuming that events of these two types in sample
space (mutations and HGT) are independent though not mutually exclusive.

Regardless of the value of $p$ or $r$, for $i=j$ we always have $\pi_{ii}=1$
(assuming $0^0=1$). As a result, every undirected graph $G$ sampled from the
random graph given by Eq.~(\ref{pij}) necessarily has self-loops at all
genotypes, no matter how sparsely interconnected the genotypes may be as a
function of $p$ and $r$.

\subsection{Network dynamics}
\label{sec:dyn}

For a fixed graph $G$ sampled from the random graph introduced in
Sec.~\ref{sec:str}, we describe the corresponding network dynamics as a set of
$2^L$ coupled differential equations, one for each possible genotype. These
equations bring together all the elements introduced in Sec.~\ref{sec:fit},
which led to the definition of the fitness $f_i=2^{d_i}$ of genotype $i$ for
$d_i$ as in Eq.~(\ref{di}), and those introduced in Sec.~\ref{sec:str}, where
the random graph giving rise to $G$ was specified via the edge probability
$\pi_{ij}$ of Eq.~(\ref{pij}) that an edge exists between genotypes $i$ and $j$.

The probabilities $p^{H_{ij}}$ and $r^{J_{ij}}$ appearing in the expression for
$\pi_{ij}$ are, as a matter of principle, expected to be relevant also to our
dynamical equations. If so, they must appear in the equations in normalized
form, as follows. Letting $N_i$ denote the set of genotypes (including $i$
itself) to which genotype $i$ is joined by an edge in graph $G$, probability
$p^{H_{ij}}$ becomes
\begin{equation}
q_{j\to i}=\frac{p^{H_{ij}}}{\sum_{k\in N_j}p^{H_{k j}}},
\label{qji}
\end{equation}
so $\sum_{i\in N_j}q_{j\to i}=1$, and similarly probability $r^{J_{ij}}$ becomes
\begin{equation}
s_{i\to j}=\frac{r^{J_{ij}}}{\sum_{k\in N_i}r^{J_{ik}}},
\label{sij}
\end{equation}
so $\sum_{j\in N_i}s_{i\to j}=1$. Probability $q_{j\to i}$ is the probability
that genotype $j$ gives rise to genotype $i$ during binary fission by undergoing
mutation at $H_{ij}$ of its nucleotides. Probability $s_{i\to j}$, in turn, is
the probability that genotype $i$ gives rise to genotype $j$ when $J_{ij}$ genes
of genotype $i$ undergo HGT.

The differential equation describing the evolution in time of the absolute
abundance of genotype $i$, $X_i$, is
\begin{equation}
\dot{X}_i=
\sum_{j\in N_i}q_{j\to i}f_jX_j+\lambda\sum_{j\in N_i}s_{i\to j}f_iX_i\\.
\label{Xip1}
\end{equation}
The first summation on the right-hand side of this equation accounts for the
total contribution to $\dot{X_i}$ from mutations affecting genotypes $j\in N_i$.
The individual contribution from each such $j$ depends on probability
$q_{j\to i}$, on the fitness of genotype $j$, and on its abundance. The second
summation is similar, now accounting for the total contribution to $\dot{X_i}$
when genes get transferred from genotype $i$ to genotypes $j\in N_i$ via HGT. In
Eq.~(\ref{Xip1}), $\lambda$ is a parameter that can be used to regulate the
relative rate at which the two types of contribution to $\dot{X}_i$ occur.
Because $\sum_{j\in N_i}s_{i\to j}=1$, Eq.~(\ref{Xip1}) can be immediately
rewritten as
\begin{equation}
\dot{X}_i=\sum_{j\in N_i}q_{j\to i}f_jX_j+\lambda f_iX_i,
\label{Xip2}
\end{equation}
where it becomes clear that the second summation on the right-hand side of
Eq.~(\ref{Xip1}) depends only on the fitness and abundance of genotype $i$.
Thus, the probabilities $r^{J_{ij}}$ giving rise to $s_{i\to j}$ through
Eq.~(\ref{sij}) are relevant only insofar as they affect the sampling of graph
$G$ from the random graph.

In spite of the presumed function of parameter $\lambda$ in Eqs.~(\ref{Xip1})
and~(\ref{Xip2}), the simplified equation for $\dot{X}_i$ given in
Eq.~(\ref{Xip2}) highlights the fact that the contribution of $f_iX_i$ to
$\dot{X}_i$ is actually multiplied by $q_{i\to i}+\lambda$. Without $\lambda$
weighing in like this, such factor would already be exponentially higher than
that of any other $f_j X_j$ for $p<1$, that is, for $j\neq i$ we would have
$q_{i\to i}/q_{j\to i}=(1/p)^{H_{ij}}$. It seems, therefore, that $\lambda$ has
no clear role to play, so henceforth we use $\lambda=0$. The final form of the
equation for $\dot{X}_i$ is then
\begin{equation}
\dot{X}_i=\sum_{j\in N_i}q_{j\to i}f_jX_j.
\label{Xi}
\end{equation}

In order to avoid the unbounded growth of $X_i$ that Eq.~(\ref{Xi}) may entail,
we rewrite it for the genotypes' relative abundances instead. Such abundances
are as given in Eq.~(\ref{xj}), i.e., the relative abundance of genotype $i$ is
$x_i=X_i/\sum_{k\in N}X_k$, so $\sum_{i\in N}x_i=1$ holds at all times. We
obtain
\begin{align}
\dot{x}_i
&=\frac{\dot{X}_i}{\sum_{k\in N}X_k}-
x_i\frac{\sum_{k\in N}\dot{X}_k}{\sum_{k\in N}X_k}\cr
&=\sum_{j\in N_i}q_{j\to i}f_jx_j-
x_i\sum_{k\in N}\sum_{j\in N_k}q_{j\to k}f_jx_j.
\end{align}
Letting
\begin{equation}
\phi=\sum_{k\in N}\sum_{j\in N_k}q_{j\to k}f_jx_j
\label{phi}
\end{equation}
yields
\begin{equation}
\dot{x}_i=\sum_{j\in N_i}q_{j\to i}f_jx_j-\phi x_i.
\label{xi}
\end{equation}
Note that, even though this equation embodies the new nonlinearities implied by
the term involving $\phi$, the truly striking ones are those already present in
Eq.~(\ref{Xi}), that is, those through which genotype fitnesses depend on
relative abundances exponentially.

\section{A special case with one single producer}
\label{sec:sc1}

By Eq.~(\ref{pij}), letting $p=r=0$ implies $\pi_{ij}=[i=j]$.\footnote{For a
logical proposition $P$, the Iverson bracket $[P]$ equals $1$ if $P$ is true,
$0$ if $P$ is false. This notation generalizes the Kronecker delta, since
$[i=j]=\delta_{ij}$.} That is, graph $G$ has no edges other than the $2^L$
self-loops, so by Eq.~(\ref{qji}) we have $q_{i\to i}=1$ for every genotype $i$.
It might seem that such a trivial topology, implying as it does that genotypes
never undergo any form of random variation, would be unable to give rise to
interesting dynamics. This is not necessarily so, however, because genotypes
still influence one another through the close coupling that their
abundance-dependent fitnesses provide.

Letting
\begin{equation}
\bar{f}=\sum_{k\in N}f_kx_k,
\end{equation}
it follows from Eq.~(\ref{phi}) that $\phi=\bar{f}$, thus leading, by
Eq.~(\ref{xi}), to
\begin{equation}
\dot{x}_i=(f_i-\bar{f})x_i.
\label{xis1}
\end{equation}
Here we look at the case in which $\mu_i=[i=0]$ (genotype $0$ is the only
producer of public goods) and $\sigma_{ji}=[j=0][i\neq 0]\sigma$ (all genotypes
other than $0$ are equally impacted by the public goods that genotype $0$
produces) for some $\sigma\in[-1,1]$. All $\sigma_{ji}$'s with $j\neq 0$ are
irrelevant. (A similar special case, but with multiple producers of public
goods, is analyzed in the Appendix.)

Assuming $x_i(0)=2^{-L}$ for every genotype $i$, Eq.~(\ref{xis1}) implies that,
in the long run, only the genotypes whose initial fitnesses are greatest
survive. If this happens for more than one genotype, then all of them survive
with the same abundance. Of all genotypes $i\neq 0$, identifying the fittest
depends on the value of $\sigma$. For $\sigma>0$, the fittest are the
$(L/B)(2^B-1)$ genotypes for which $J_{0i}=1$. For $\sigma=0$, all $2^L-1$
genotypes have the same fitness. For $\sigma<0$, the fittest are the
$(2^B-1)^{L/B}$ genotypes for which $J_{0i}=L/B$. It follows that survival is
determined by how $\sigma_{00}$ relates to $\sigma u$ (if $\sigma_{00}>0$ and
$\sigma>0$), or to $\sigma u^{L/B}$ (if $\sigma_{00}<0$ and $\sigma<0$), or
directly to $\sigma$ (in all other cases). In general, we denote the greatest
fitness among genotypes $i\neq 0$ by $f_+$. There are nine cases to be
considered, discussed next. In this discussion, any conclusion that a group of
genotypes $i\neq 0$ survives is to be extended to all genotypes $i\neq 0$ if
$u=1$, since in this case they all have the same fitness.
\begin{enumerate}
\item[C1.]
$\sigma_{00}>0$ and $\sigma>0$:
$f_0>1$ and $f_+>1$, with three sub-cases:
\begin{itemize}
\item[C1.a.]
$\sigma_{00}>\sigma u$:
$f_0>f_+$;
genotype $0$ survives.
\item[C1.b.]
$\sigma_{00}=\sigma u$:
$f_0=f_+$;
genotype $0$ survives, and so do all genotypes $i$ such that $J_{0i}=1$.
\item[C1.c.]
$\sigma_{00}<\sigma u$:
$f_0<f_+$;
genotypes $i$ such that $J_{0i}=1$ survive.
\end{itemize}
\item[C2.]
$\sigma_{00}>0$ and $\sigma=0$:
$f_0>1$ and $f_+=1$;
genotype $0$ survives.
\item[C3.]$\sigma_{00}>0$ and $\sigma<0$:
$f_0>1$ and $f_+<1$;
genotype $0$ survives.
\item[C4.]
$\sigma_{00}=0$ and $\sigma>0$:
$f_0=1$ and $f_+>1$;
genotypes $i$ such that $J_{0i}=1$ survive.
\item[C5.]
$\sigma_{00}=0$ and $\sigma=0$:
$f_0=1$ and $f_+=1$;
all genotypes survive.
\item[C6.]
$\sigma_{00}=0$ and $\sigma<0$:
$f_0=1$ and $f_+<1$;
genotype $0$ survives.
\item[C7.]
$\sigma_{00}<0$ and $\sigma>0$:
$f_0<1$ and $f_+>1$;
genotypes $i$ such that $J_{0i}=1$ survive.
\item[C8.]
$\sigma_{00}<0$ and $\sigma=0$:
$f_0<1$ and $f_+=1$;
genotypes $i\neq 0$ survive.
\item[C9.]
$\sigma_{00}<0$ and $\sigma<0$:
$f_0<1$ and $f_+<1$, with three sub-cases:
\begin{itemize}
\item[C9.a.]
$\sigma_{00}>\sigma u^{L/B}$:
$f_0>f_+$;
genotype $0$ survives.
\item[C9.b.]
$\sigma_{00}=\sigma u^{L/B}$:
$f_0=f_+$;
genotype $0$ survives, and so do all genotypes $i$ such that $J_{0i}=L/B$.
\item[C9.c.]
$\sigma_{00}<\sigma u^{L/B}$:
$f_0<f_+$;
genotypes $i$ such that $J_{0i}=L/B$ survive.
\end{itemize}
\end{enumerate}

A careful examination of these cases reveals that a number of outcomes are
possible in the long run, which seems remarkable as we consider that the only
edges in graph $G$ are the self-loops at all genotypes. Interestingly, these
results provide useful insight even when lifting the $p=r=0$ simplification, as
we show in Sec.~\ref{sec:res}.

\section{Results}
\label{sec:res}

All our computational results come from time stepping Eq.~(\ref{xi}) from
$x_i(0)=2^{-L}$ for all genotypes $i\in N$. This is done on a fixed graph $G$,
obtained by Monte Carlo sampling as explained in Sec.~\ref{sec:str}. The
expected steady-state value of each $x_i$ is obtained by averaging over several
such graphs.

A potentially problematic aspect of time stepping Eq.~(\ref{xi}), depending on
the case at hand, is the often very small absolute value of each of the terms
contributing to $d_i$ in Eq.~(\ref{di}). An issue to be considered is that
several genotypes may have nearly neutral (close to $1$) fitnesses for a long
time even if eventually they are to diverge from one another substantially. In
this case, convergence to the steady-state relative abundances can be slowed
down significantly. Another issue is the effect of the unavoidable round-off
errors that accompany operations on numbers very close to $0$. Because of such
errors, the value obtained for $d_i$ may depend on the order used for the terms
to enter the sum, which can break important symmetries whenever the equations
dictate that $i$ and $j$ exist for which we should have $d_i=d_j$. To avoid
this, at every step the terms that make up $d_i$ are added in increasing order
of absolute value. Thus, except for cases involving one single producer, a
sorting operation is required at each step, thus making convergence to the
steady state very slow indeed.

We sidestep these issues by reducing the size and variety of the scenarios we
consider while at the same time retaining the possibility of interesting and
diverse behavior. We thus consider only a small number of genotypes: we use
$L=9$ and $B=3$, hence $512$ genotypes, each with three $3$-nucleotide genes.
Moreover, all scenarios we consider generalize the special case of
Sec.~\ref{sec:sc1} by allowing $p,r>0$ and therefore the appearance of
nontrivial topologies (that is, those on which manifestations of the mutational
and HGT-related aspects of the model can occur). All other simplifications of
that section continue to be assumed, so $p=r$, genotype $0$ is the only producer
of public goods, and $\sigma_{0i}=\sigma$ for every genotype $i\neq 0$. Given
these simplifications, the expected behavior of all
\begin{equation}
n_J=\binom{L/B}{J}(2^B-1)^J
\end{equation}
genotypes that differ from genotype $0$ at $J$ genes is the same. A handy
simplification when analyzing results is then to consider all such genotypes as
a single group, that is, consider them through their mean relative abundances,
denoted by $x_J$ and given by
\begin{equation}
x_J=n_J^{-1}\sum_{i\in N\mid J_{0i}=J}x_i.
\label{xJ}
\end{equation}
Of course, $\sum_{J=0}^{L/B}n_Jx_J=1$.

Given the random graph used to represent our network of genotypes, the expected
degree of a randomly chosen genotype $i$ (the expected number of edges incident
to it, including its self-loop) can be obtained by summing up $\pi_{ij}$ (the
probability that an edge exists between genotypes $i$ and $j$) over $j\in N$.
Based on Eq.~(\ref{pij}), we write this summation in terms of the number $H$ of
nucleotides at which $i$ and $j$ may differ and likewise the number $J$ of
genes at which they may differ. Letting $z$ denote the desired expected degree,
we obtain
\begin{align}
z
={}&\sum_{H=0}^L\binom{L}{H}p^H+\sum_{J=0}^{L/B}n_Jr^J
-\sum_{J=0}^{L/B}\binom{L/B}{J}r^J\cr
&\sum_{n_1=1}^B\cdots\sum_{n_J=1}^B
\binom{B}{n_1}\cdots\binom{B}{n_J}p^{n_1+\cdots+n_J}\cr
={}&(1+p)^L+[1+(2^B-1)r]^{L/B}\cr
&-[1+[(1+p)^B-1]r]^{L/B}.
\end{align}
Clearly, increasing $p$ (or $r$, should it be allowed to change independently of
$p$) leads $z$ to grow as well, and with it the expected number of genotypes
that can be reached from genotype $i$ through some undirected path in the graph.

\begin{figure}[t]
\includegraphics[scale=0.35]{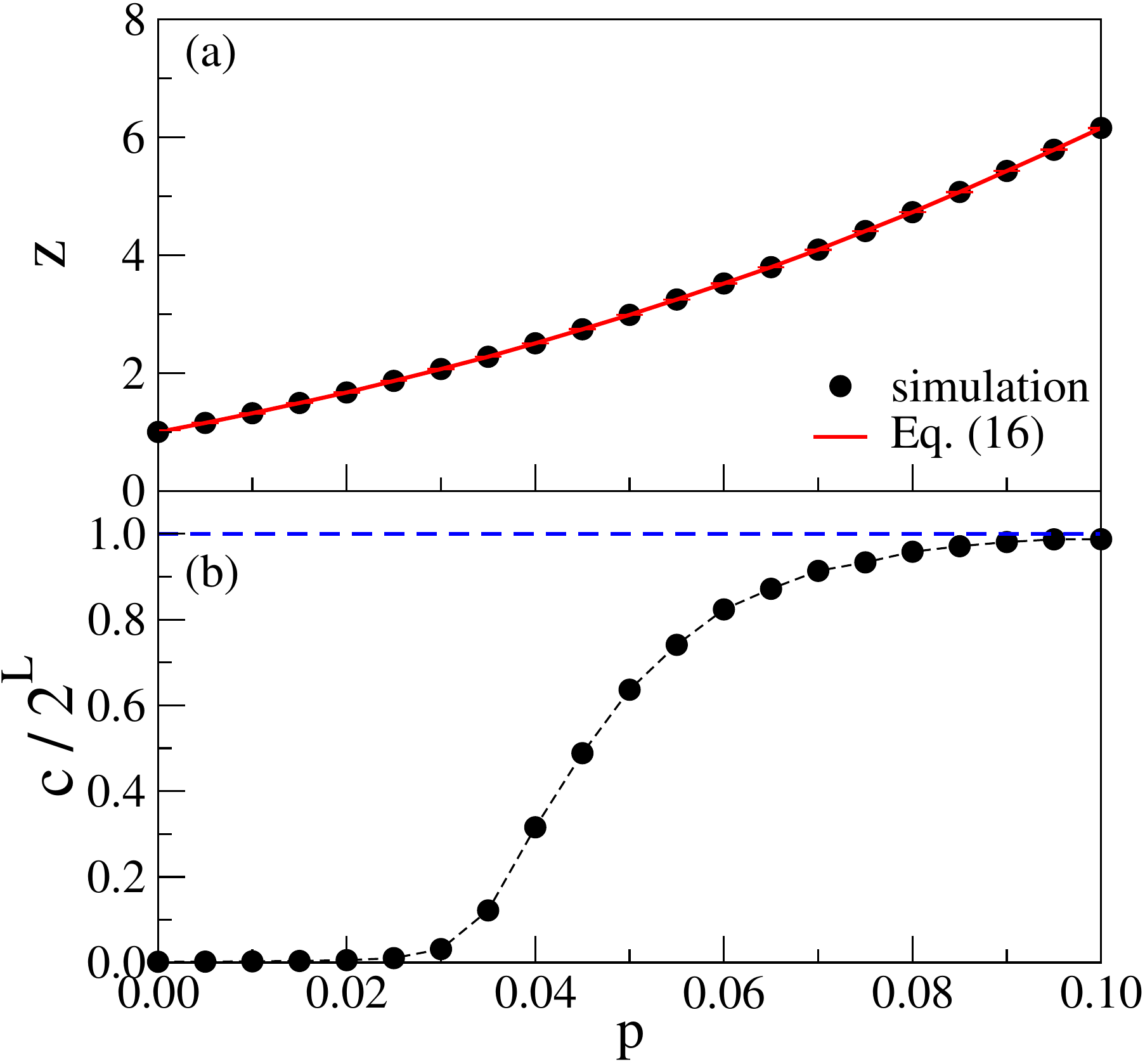}
\caption{(a) Expected degree of a randomly chosen genotype. (b) Expected
fraction of $2^L$ corresponding to the connected component to which that
genotype belongs. Both panels refer to $L=9$ and $B=3$.}
\label{fig1}
\end{figure}

Such genotypes constitute the so-called connected component to which $i$
belongs. In the context at hand, this is the set of genotypes expected to relate
to $i$ via mutation or HGT. For $i=0$, this means that probability $p$ can be
used to control the expected extent to which the behavior of the single producer
is evolutionarily related to those of other genotypes. That is, while all $2^L$
genotypes partake of the public goods produced by genotype $0$, the set of those
that interact with the producer, directly or indirectly through mutation or HGT,
can be shrunk or enlarged by controlling the value of $p$. This is illustrated
in Fig.~\ref{fig1}, where the expected number of genotypes in the connected
component is denoted by $c$. Clearly, letting $p\le 0.1$ suffices for all
interesting cases of the connected component in question to be encompassed. This
includes the case in which it only contains genotype $0$, the case in which it
contains a sizable fraction of all $2^L$ genotypes, and the transition
in-between as $p$ grows from $p=0$. All values of $p$ are thus constrained
henceforth. (Incidentally, in this case it can be verified that relaxing the
assumption $p=r$ only mildly, e.g., $0.95p\le r\le 1.05p$, causes $z$ to vary
mildly as well, about $\pm 5\%$ for $p=0.1$, less for smaller $p$. This may help
regard the $p=r$ constraint as not so stringent after all.) 

We present our results in two groups of figures, the first relating to case~C1,
of Sec.~\ref{sec:sc1}, with both $\sigma_{00}>0$ and $\sigma>0$, the second
relating to case~C9, with both $\sigma_{00}<0$ and $\sigma<0$. Note that, in
regard to cases~C1--C9 of that section, C1 and C9 are the only nontrivial ones
in that outcomes depend on how parameters other than $\sigma_{00}$ and $\sigma$
intervene. With only two exceptions, to be discussed shortly, all figures show
the steady-state value of $x_J$ for $J\in\{0,1,2,3\}$. Note, in these figures,
that because of their definition in Eq.~(\ref{xJ}) these four values of $x_J$ do
not in general add up to $1$.

All figures in the first group, Figs.~\ref{fig2}--\ref{fig5}, are for $u=0.01$
and $\sigma=1$. Figures~\ref{fig2} and~\ref{fig3} are for $p=0.075$, leading to
$z\approx 4.41$ and $c\approx 478$. Figure~\ref{fig2} is one of the exceptions
alluded to above, since it shows the steady-state $x_i$'s individually. A
similar breakdown is of course possible in all other cases of this group and the
second, but is omitted in most of them. Figures~\ref{fig4} and~\ref{fig5}
complete the first group, respectively with $p=0.045$ ($z\approx 2.74$ and
$c\approx 250$) and $p=0.0384$ ($z\approx 2.43$ and $c\approx 136$).

All figures in the second group, Figs.~\ref{fig6}--\ref{fig9}, are for $u=0.1$
and $\sigma=-1$. They refer to the same values of $p$ as the figures in the
first group, in the same order. Figure~\ref{fig6} is the second exception
alluded to earlier, since it pairs with Fig.~\ref{fig7} in that the former
refers to individual relative abundances and the latter to mean relative
abundances for the same value of $p$. Note that the value of $u$ for all figures
in this second group is greater than that of Figs.~\ref{fig2}--\ref{fig5} by
one order of magnitude. This is meant to avoid aggravating the first convergence
issue mentioned at the beginning of this section even further. As discussed in
Sec.~\ref{sec:dis}, the critical value of $J_{0i}$ for
Figs.~\ref{fig2}--\ref{fig5} is $J_{0i}=1$, whereas for
Figs.~\ref{fig6}--\ref{fig9} it is $J_{0i}=L/B=3$, so values of $\sigma_{00}$
closer to $0$ are needed.

\section{Discussion}
\label{sec:dis}

The public goods produced by bacteria in QS can be greatly diverse, including
for example enzymes \cite{wdg17} and vitamins \cite{sst20}, and are therefore
essential for a number of cellular functions related to growth and metabolism.
However, a cell's production of public goods has costs associated with it that
have the potential to completely offset the benefits accrued to the cell itself
by autoinduction. A central factor affecting the tipping of this balance one way
or the other is the presence of certain mutants, commonly referred to as
``cheaters,'' that do not join in the production of public goods but may
nevertheless benefit from them. Whether the presence of cheater genotypes can
negatively impact the survival of genotypes that do produce public goods depends
on a variety of issues. Ultimately, these issues can be summarized as two
overall strategies, one highlighting positive effects on producers even if
cheaters might be positively affected as well, the other highlighting negative
effects on cheaters even if producers might be negatively affected as well. The
first (Strategy~1) is autoregulation of the costs versus benefits of public-good
production, that is, public goods are only produced if the latter outweigh the
former (cf., e.g., \cite{xkf11}). The second (Strategy~2) is the exploitation of
some weakening effect associated with public-good uptake by cheaters (e.g.,
\cite{dcg12,msg16}). All our computational results have genotype $0$ as the sole
producer of public goods and are therefore focused on the trade-offs affecting
the survival of genotype $0$ and all $2^L-1$ cheaters.

\begin{figure}[t]
\includegraphics[scale=0.35]{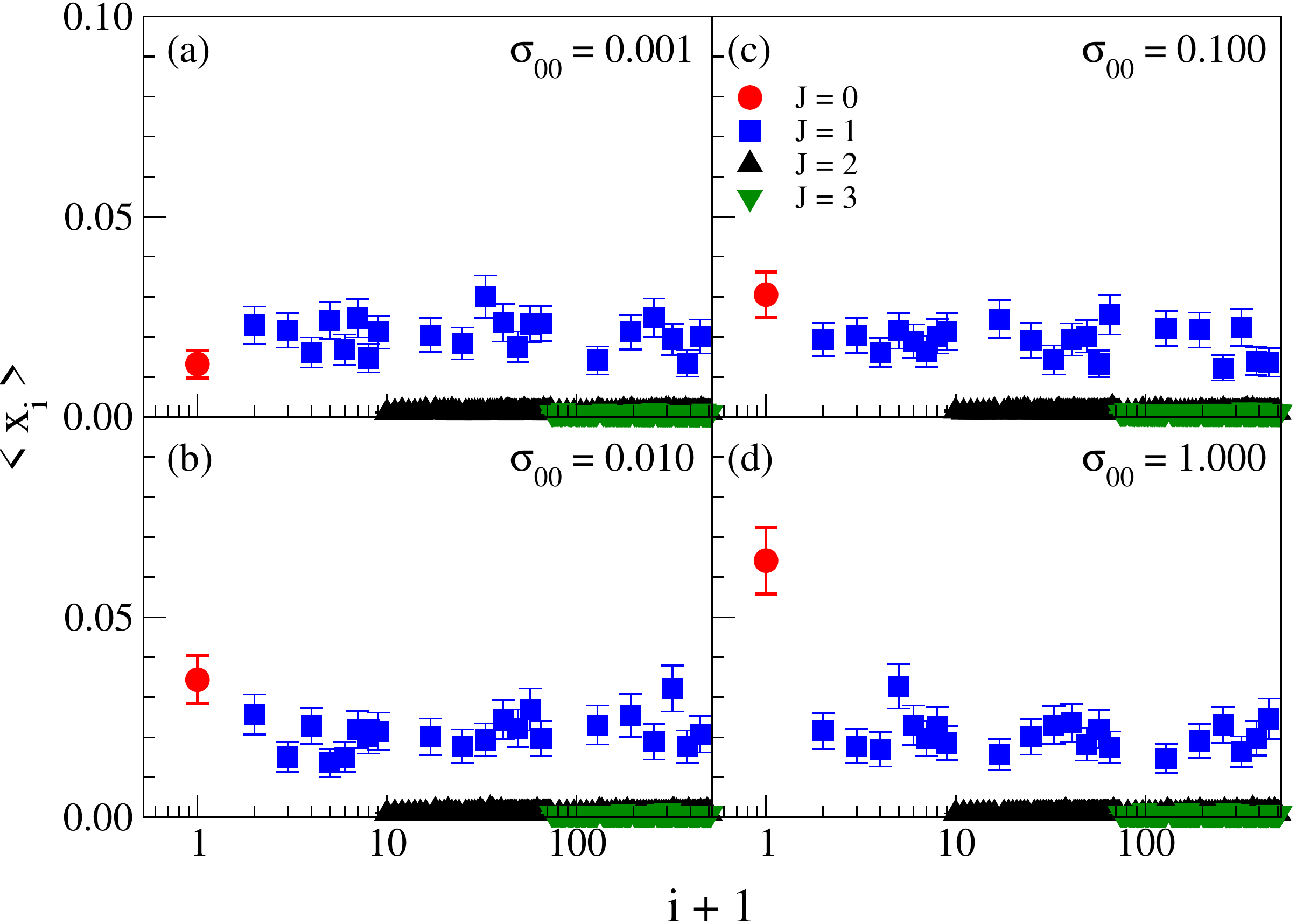}
\caption{Average steady-state individual relative abundances ($x_i$) for $L=9$,
$B=3$, $p=0.075$ ($z\approx 4.41$, $c\approx 478$), $u=0.01$, and $\sigma=1$.
Values of $\sigma_{00}$ are $0.001$ (a), $0.01$ (b), $0.1$ (c), and $1$ (d). See
Fig.~\ref{fig3} for the corresponding mean relative abundances ($x_J$).}
\label{fig2}
\end{figure}

\begin{figure}[t]
\includegraphics[scale=0.35]{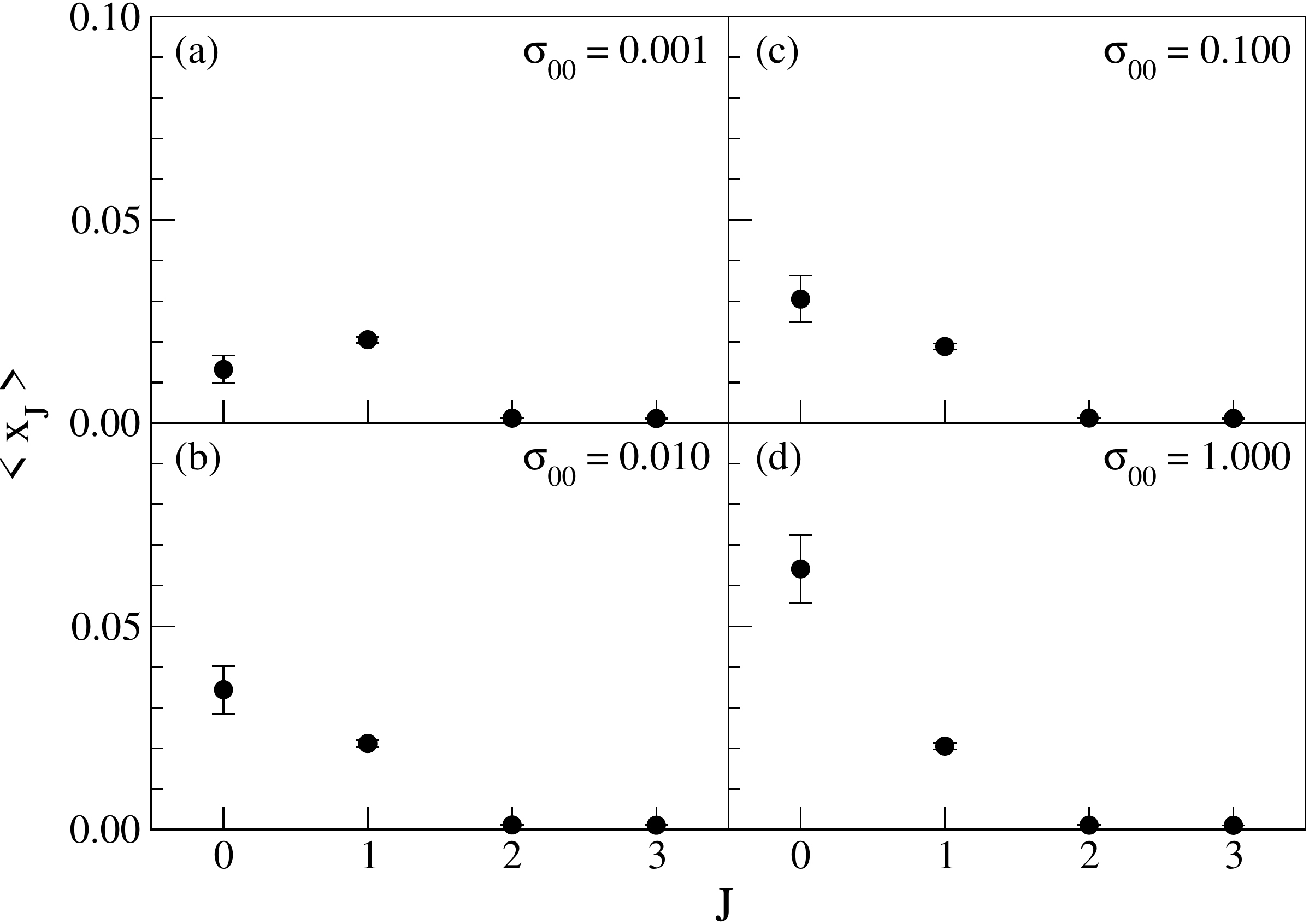}
\caption{Average steady-state mean relative abundances ($x_J$) for $L=9$, $B=3$,
$p=0.075$ ($z\approx 4.41$, $c\approx 478$), $u=0.01$, and $\sigma=1$. Values of
$\sigma_{00}$ are $0.001$ (a), $0.01$ (b), $0.1$ (c), and $1$ (d). See
Fig.~\ref{fig2} for the corresponding individual relative abundances ($x_i$).}
\label{fig3}
\end{figure}

\begin{figure}[t]
\includegraphics[scale=0.35]{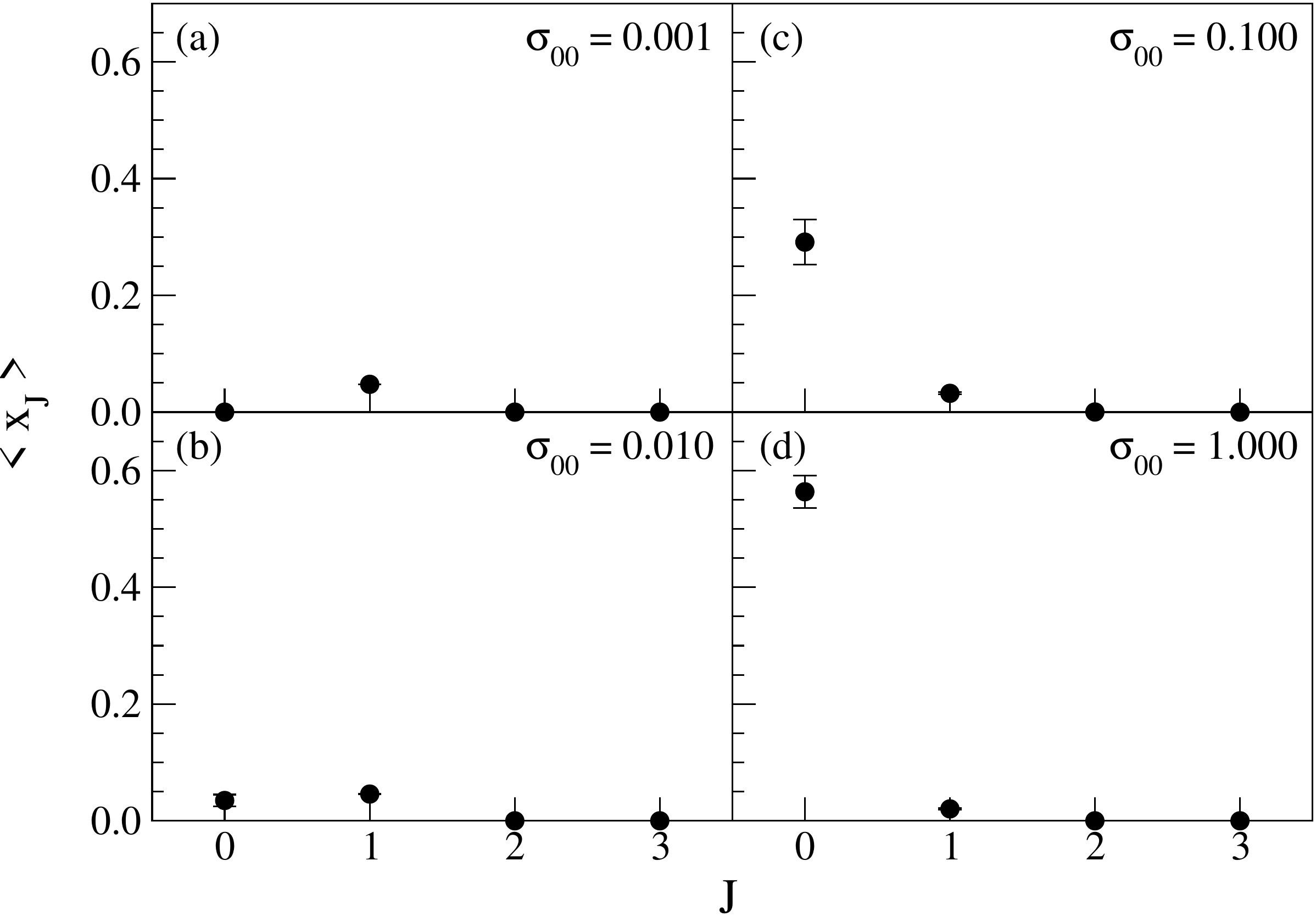}
\caption{Average steady-state mean relative abundances ($x_J$) for $L=9$, $B=3$,
$p=0.045$ ($z\approx 2.74$, $c\approx 250$), $u=0.01$, and $\sigma=1$. Values of
$\sigma_{00}$ are $0.001$ (a), $0.01$ (b), $0.1$ (c), and $1$ (d).}
\label{fig4}
\end{figure}

\begin{figure}[t]
\includegraphics[scale=0.35]{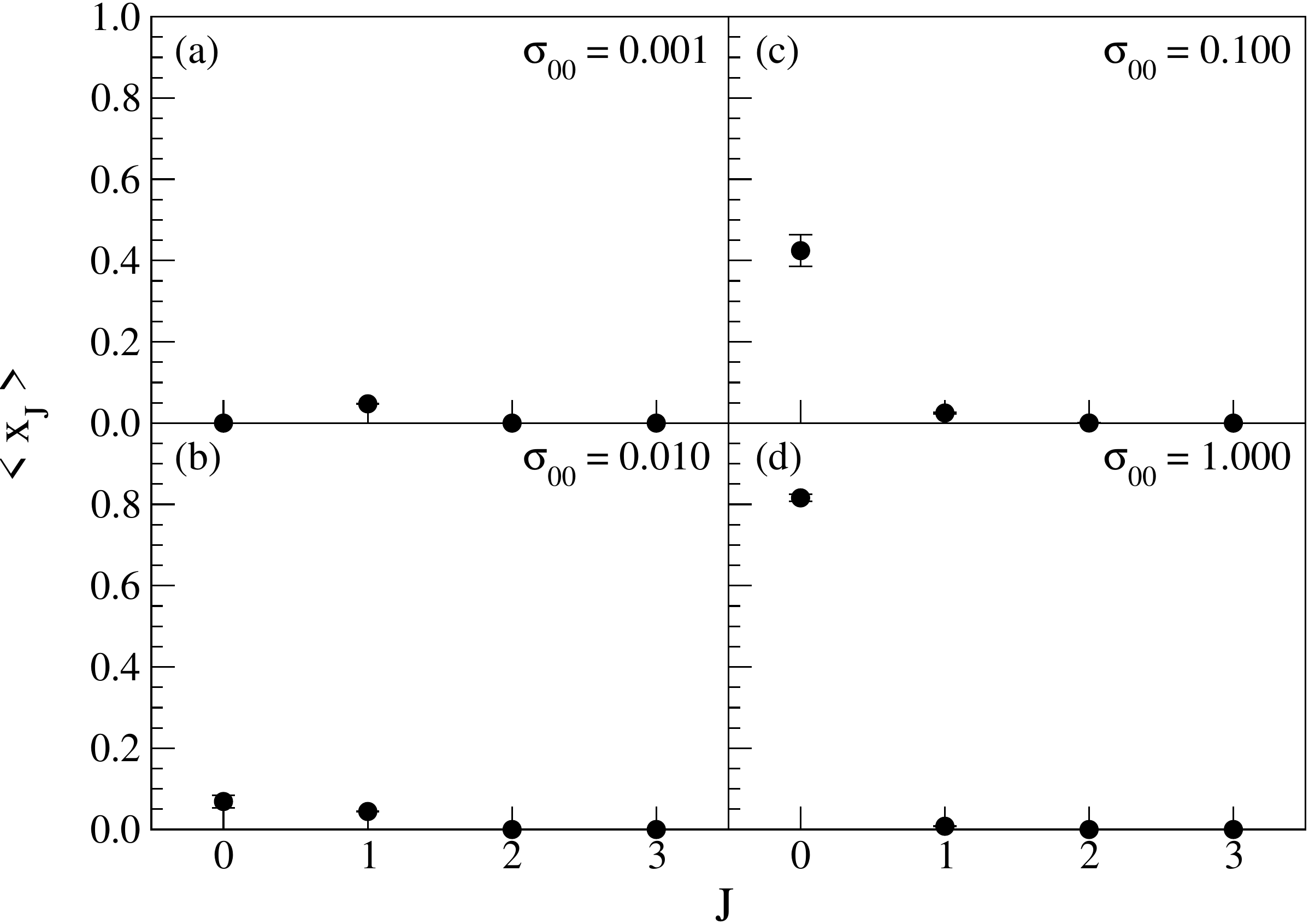}
\caption{Average steady-state mean relative abundances ($x_J$) for $L=9$, $B=3$,
$p=0.0384$ ($z\approx 2.43$, $c\approx 136$), $u=0.01$, and $\sigma=1$. Values
of $\sigma_{00}$ are $0.001$ (a), $0.01$ (b), $0.1$ (c), and $1$ (d).}
\label{fig5}
\end{figure}

The first group of results (depicted in Figs.~\ref{fig2}--\ref{fig5}) is for
$\sigma_{00},\sigma>0$ and can be seen as modeling Strategy~1, since
$\sigma_{00}>0$ can be taken as indicating that for genotype $0$ public-good
benefits outweigh production costs, and $\sigma>0$ as indicating that cheaters
benefit from the public goods without significant hindrance. These results are
for $\sigma u=0.01$, so in the absence of any mutational or HGT-related exchange
between genotypes (as in Sec.~\ref{sec:sc1}, case~C1) we would have a sharp
threshold for $\sigma_{00}$ at $0.01$ separating the exclusive survival of
genotype $0$ (for $\sigma_{00}$ above the threshold) from the exclusive survival
of the $n_1=21$ genotypes $i$ having $J_{0i}=1$ (for $\sigma_{00}$ below the
threshold). For $\sigma_{00}$ precisely at the threshold the two groups would
coexist.

Figures~\ref{fig2} and~\ref{fig3} both refer to scenarios in which, on average,
genotype $0$ undergoes genetic interactions, directly or indirectly, with about
$(478-1)/(512-1)\approx 93.3\%$ of all other genotypes. Figure~\ref{fig2} shows
individual relative abundances in the long run, Fig.~\ref{fig3} mean relative
abundances. The value of $\sigma_{00}$ is varied over three orders of magnitude
from panel~(a) through panel~(d) in each figure, leading genotype $0$ from a
situation of very low relative abundance for $\sigma_{00}=0.001$ to one of clear
preponderance for $\sigma_{00}=1$. All along the competition is seen to be
taking place between genotype $0$ ($J=0$) and those that differ from it at
exactly $J=1$ gene. All other genotypes are seen to be heading toward perishing.
This progression along the increasing values of $\sigma_{00}$ seems to lead to
something resembling the outcome for case~C1 of Sec.~\ref{sec:sc1}, but there
are important differences as well, all owing to the possibility of genetic
interactions between genotype $0$ and the others. One of these differences is
that the $J=1$ genotypes never vanish from the scene, not even as genotype $0$
becomes the most abundant one. Another is that the sharp threshold at
$\sigma_{00}=0.01$ separating the demise of genotype $0$ from the other regimes
in Sec.~\ref{sec:sc1} now seems spread over at least two orders of magnitude.

Figures~\ref{fig4} and~\ref{fig5} show the effect of restricting the
possibilities of genetic interaction with genotype $0$, first to about
$(250-1)/(512-1)\approx 48.7\%$ of all other genotypes (Fig.~\ref{fig4}), then
to $(136-1)/(512-1)\approx 26.4\%$ (Fig.~\ref{fig5}). While the overall trends
relating the survival of genotype $0$ to that of the $J=1$ genotypes remain the
same, it seems clear that genotype $0$ comes ever closer to being the sole
survivor as $\sigma_{00}$ is increased and its possibilities of genetic
interaction become limited to a smaller neighborhood in the graph.

\begin{figure}[t]
\includegraphics[scale=0.35]{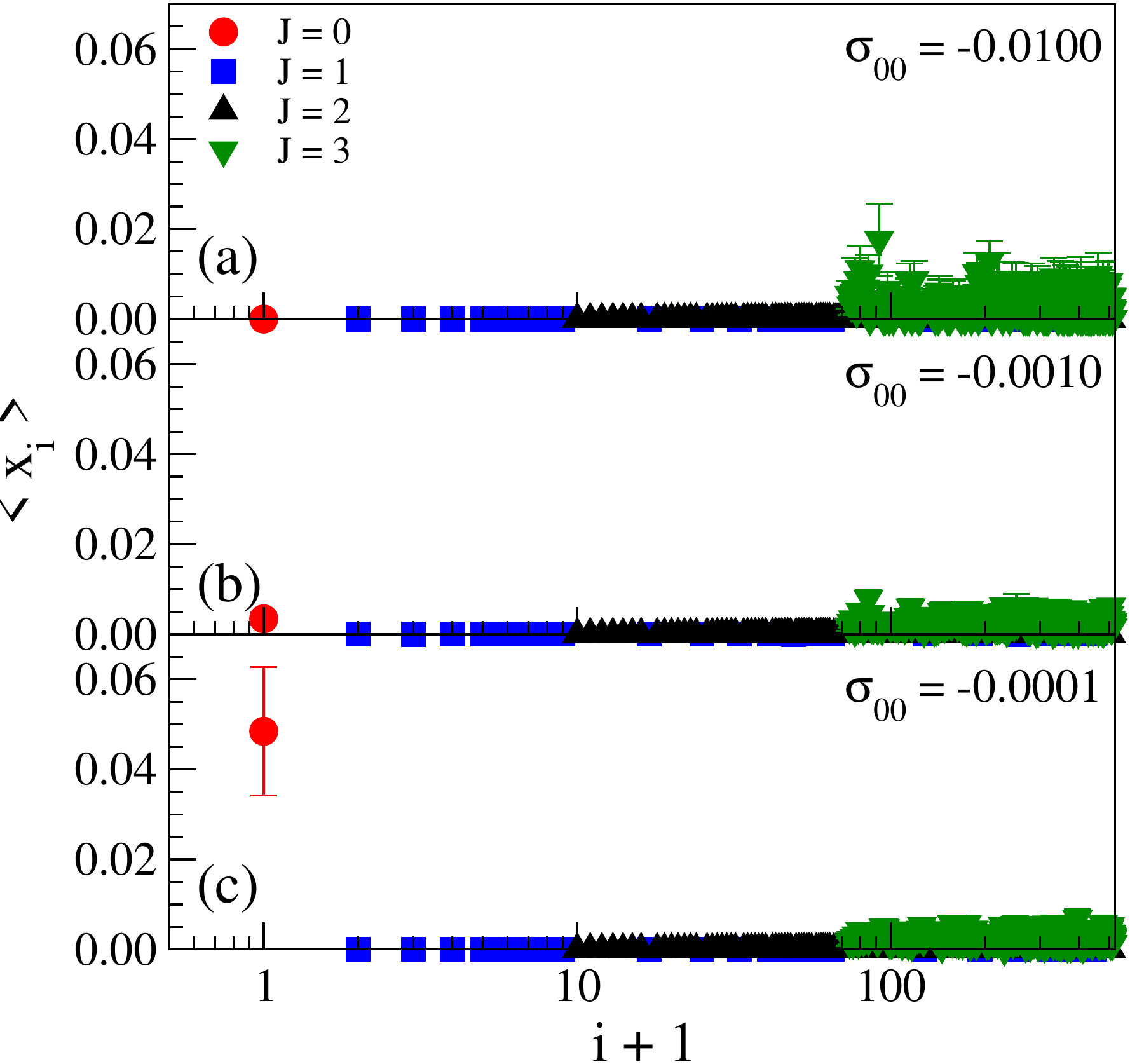}
\caption{Average steady-state individual relative abundances ($x_i$) for $L=9$,
$B=3$, $p=0.075$ ($z\approx 4.41$, $c\approx 478$), $u=0.1$, and $\sigma=-1$.
Values of $\sigma_{00}$ are $-0.01$ (a), $-0.001$ (b), and $-0.0001$ (c). See
Fig.~\ref{fig7} for the corresponding mean relative abundances ($x_J$).}
\label{fig6}
\end{figure}

\begin{figure}[t]
\includegraphics[scale=0.35]{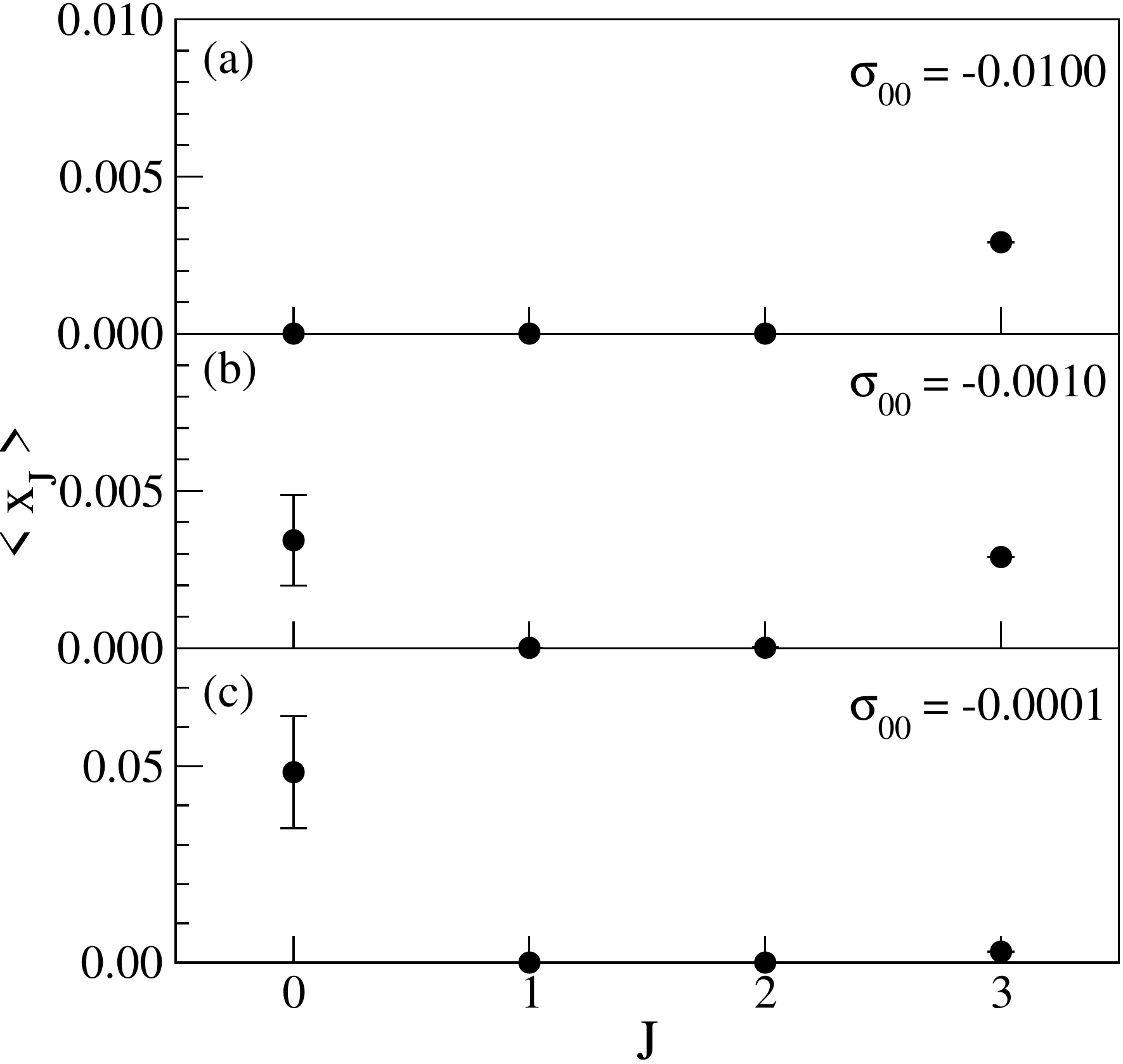}
\caption{Average steady-state mean relative abundances ($x_J$) for $L=9$, $B=3$,
$p=0.075$ ($z\approx 4.41$, $c\approx 478$), $u=0.1$, and $\sigma=-1$. Values of
$\sigma_{00}$ are $-0.01$ (a), $-0.001$ (b), and $-0.0001$ (c). Note the
different ordinate scales. See Fig.~\ref{fig6} for the corresponding
individual relative abundances ($x_i$).}
\label{fig7}
\end{figure}

\begin{figure}[t]
\includegraphics[scale=0.35]{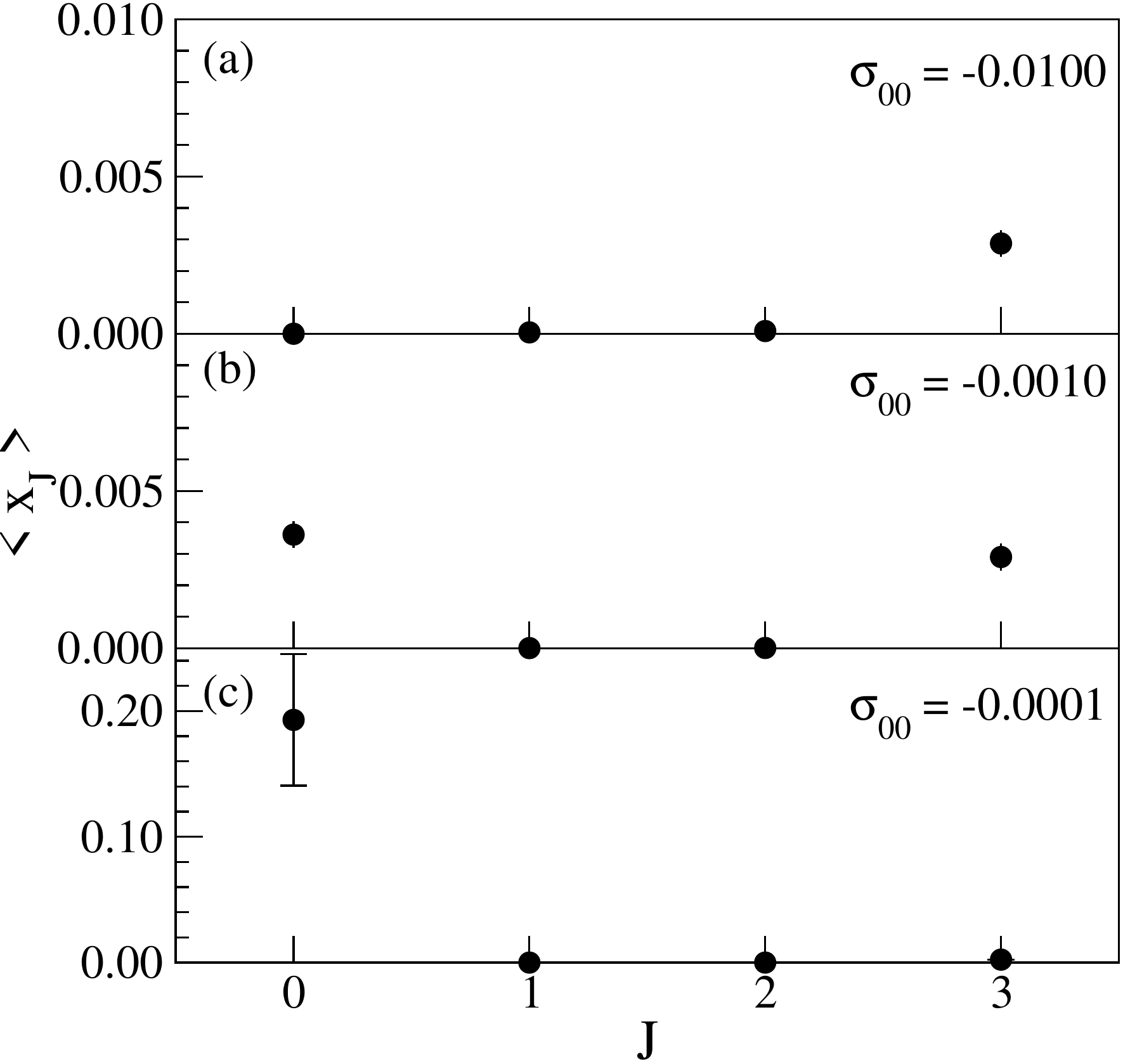}
\caption{Average steady-state mean relative abundances ($x_J$) for $L=9$, $B=3$,
$p=0.045$ ($z\approx 2.74$, $c\approx 250$), $u=0.1$, and $\sigma=-1$. Values of
$\sigma_{00}$ are $-0.01$ (a), $-0.001$ (b), and $-0.0001$ (c). Note the
different ordinate scales.}
\label{fig8}
\end{figure}

\begin{figure}[t]
\includegraphics[scale=0.35]{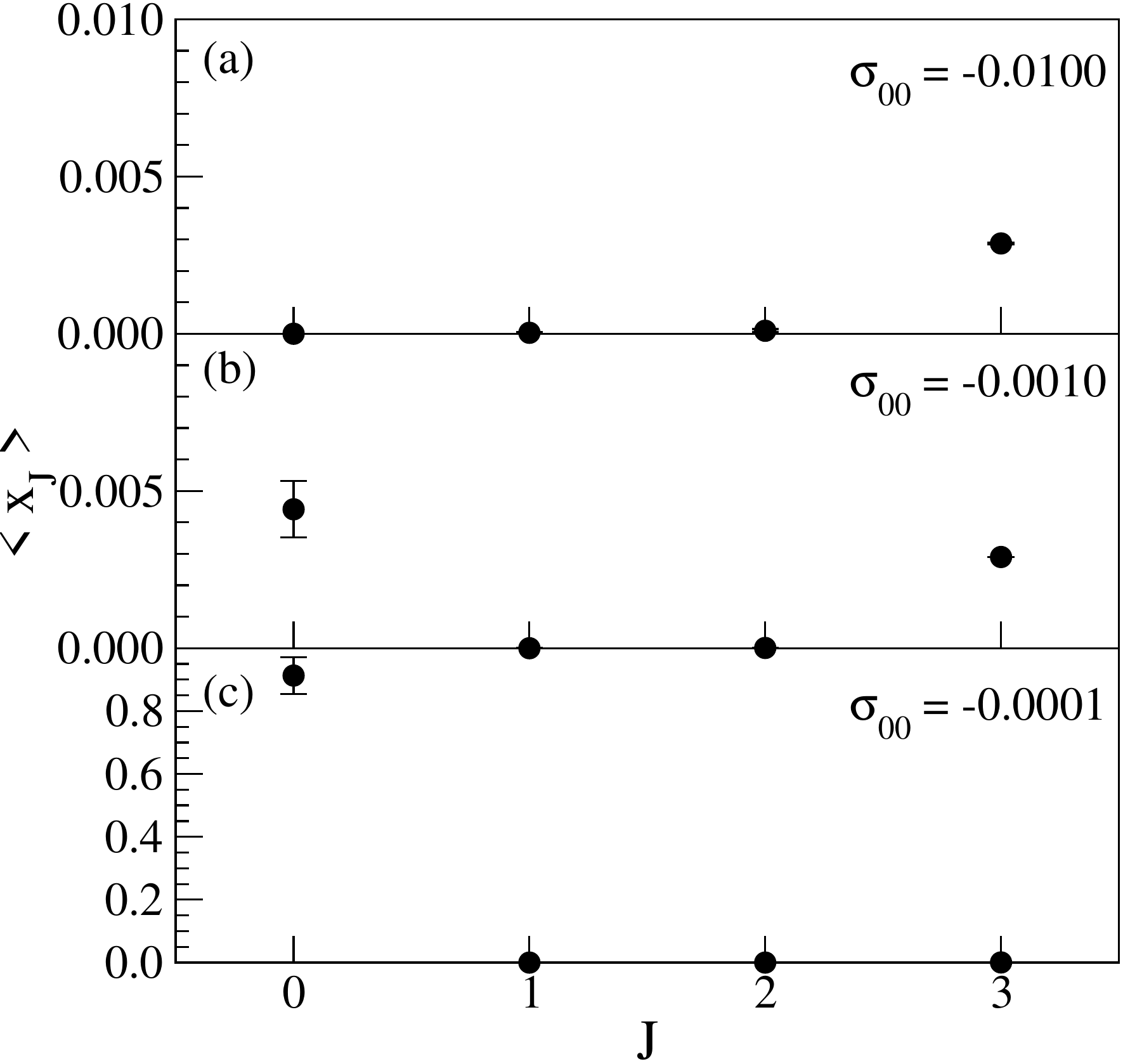}
\caption{Average steady-state mean relative abundances ($x_J$) for $L=9$, $B=3$,
$p=0.0384$ ($z\approx 2.43$, $c\approx 136$), $u=0.1$, and $\sigma=-1$. Values
of $\sigma_{00}$ are $-0.01$ (a), $-0.001$ (b), and $-0.0001$ (c). Note the
different ordinate scales.}
\label{fig9}
\end{figure}

The second group of results (depicted in Figs.~\ref{fig6}--\ref{fig9}) is for
$\sigma_{00},\sigma<0$. This group can be regarded as modeling Strategy~2, with
$\sigma_{00}<0$ reflecting the downside of the absence of autoregulation
(public-good benefits are outweighed by production costs) and $\sigma<0$
indicating that cheaters pay a heavy price for absorbing potentially beneficial
public goods without joining in producing them. Results are now for
$\sigma u^{L/B}=-0.001$, which like before would constitute a sharp threshold
for $\sigma_{00}$ at $-0.001$ should genotypes undergo no mutational or
HGT-related interactions (as in Sec.~\ref{sec:sc1}, case~C9). Now the divide
would be between the exclusive survival of genotype $0$ (for $\sigma_{00}$ above
the threshold) and the exclusive survival of the $n_3=343$ genotypes $i$ having
$J_{i0}=3$ (for $\sigma_{00}$ below the threshold). Coexistence would ensue for
$\sigma_{00}$ precisely at the threshold.

In each of Figs~\ref{fig6}--\ref{fig9} the value of $\sigma_{00}$ is increased
by one order of magnitude from panel~(a) to panel~(b), then by one more from
panel~(b) to panel~(c). Along these increases the competition for survival takes
place between genotype $0$ and those that differ from it at exactly $J=3$ genes.
All other genotypes are heading toward perishing. In all four figures the
genotypes having $J=3$ survive alone for $\sigma_{00}=-0.01$. For larger values
of $\sigma_{00}$ genotype $0$ is seen to survive as well, becoming the most
abundant one for $\sigma_{00}=-0.0001$. As $\sigma_{00}$ is thus increased a
transition similar to that of case~C9 of Sec.~\ref{sec:sc1} takes place, but
once again this happens much less abruptly, along at least one order of
magnitude.

As with the first group of figures, moving from Figs.~\ref{fig6} and~\ref{fig7}
to Fig.~\ref{fig8} and then to Fig.~\ref{fig9} allows us to track the effects of
letting genotype $0$ interact genetically with progressively fewer other
genotypes, from about $93.3\%$ of them in Figs.~\ref{fig6} and~\ref{fig7}, to
about $48,7\%$ in Fig.~\ref{fig8}, then about $26.4\%$ in Fig.~\ref{fig9}.
Clearly, as genotype $0$ becomes more confined in its possibilities for genetic
interaction, so does its prevalence relative to the $J=3$ genotypes become more
pronounced.

\section{Conclusion}
\label{sec:con}

We have studied the evolutionary dynamics of bacterial QS when multiple species
(here referred to as genotypes) participate. We have focused on the production
and uptake of public goods by the cells and on the trade-offs arising when
cheater genotypes (those that have the potential to benefit from the uptake of
public goods but do not join in producing them) are present. Our model has a
number of parameters intended to allow for several scenarios to be considered.
Two probability parameters ($p,r$) are related to how genotypes interact
genetically, and another ($u$) is related to how molecular compatibility between
genotypes influences the uptake by one genotype of public goods produced by
another. Further parameters specify which genotypes are producers (the
$\mu_i$'s) and whether (and how strongly) public-good uptake is detrimental or
beneficial to a genotype's fitness (the $\sigma_{ji}$'s).

All our computational results refer to genotype $0$ being the sole producer, all
others being cheaters. All cheaters would be equally affected by the uptake of
the public goods produced by genotype $0$ if this only depended on the
$\sigma_{0i}$'s (i.e., all $\sigma_{0i}$'s equal some fixed $\sigma$), but the
action of parameter $u$ causes significant differentiation. For fixed genotype
and gene sizes, we have found that the long-term survival of genotype $0$
depends chiefly on how $\sigma_{00}$ stands relative to a function of $\sigma$
and $u$ (though parameters $p$ and $r$ also have a part to play). Our analytical
results in the Appendix suggest that a similar conclusion may come to hold when
multiple producers are considered.

Our model can be easily adapted, mainly by altering the modes of genetic
interaction between genotypes, to the study of QS in other microorganisms (e.g.,
fungi \cite{hjltjsdn01}). In fact, to varying degrees the molecule-mediated
communication that characterizes QS in microorganisms is observed also in other
biological systems, from simple infectious agents such as viruses
\cite{essdspmlsaas17}, to communities of so-called artificial cells
\cite{beh20}, to precursor-cell clusters as they differentiate into specific
organs and tissues (sometimes with surprisingly accurate global results; cf.,
e.g., \cite{aabhbb11}). We expect that also in some of these systems the
analysis of long-term global behavior could benefit from the development of
models closely related to the one we have presented.

\begin{acknowledgments}
We acknowledge partial support from Conselho Nacional de Desenvolvimento
Cient\'\i fico e Tecnol\'ogico (CNPq), Coordena\c c\~ao de Aperfei\c coamento de
Pessoal de N\'\i vel Superior (CAPES), and a BBP grant from Funda\c c\~ao Carlos
Chagas Filho de Amparo \`a Pesquisa do Estado do Rio de Janeiro (FAPERJ).
We also acknowledge support from Agencia Nacional de Investigaci\'on e
Innovaci\'on (ANII) and Programa de Desarrollo de las Ciencias B\'asicas
(PEDEClBA). We thank N\'ucleo Avan\c cado de Computa\c c\~ao de Alto Desempenho
(NACAD), Instituto Alberto Luiz Coimbra de P\'os-Gradua\c c\~ao e Pesquisa em
Engenharia (COPPE), Universidade Federal do Rio de Janeiro (UFRJ), for the use
of supercomputer Lobo Carneiro, where most of the calculations were carried out.
\end{acknowledgments}

\appendix*
\section{A special case with multiple producers}

The special case of Sec.~\ref{sec:sc1} can be extended to allow multiple
producers of public goods while remaining tractable as far as predicting
behavior in the long run is concerned. While studying multiple-producer
scenarios computationally has proven infeasible given the resources we can
presently access, the special case we analyze in this appendix serves to
illustrate a more complex situation in which the fate of producers and certain
other genotypes continues to be determined, in the nontrivial cases, by a
comparison between some producer $k$'s $\sigma_{kk}$ and a threshold.

The extension of Sec.~\ref{sec:sc1} we now present preserves the $p=r=0$
assumption of that section, so graph $G$ continues to have only self-loops for
edges and the dynamics continues to be given by Eq.~(\ref{xis1}). For uniform
initial abundances, which we continue to assume, survival in the long run
continues to occur for the initially fittest genotypes only, always with the
same abundance.

Let $P$ be the set of producers, i.e., $\mu_i=[i\in P]$. Assume $L/B\ge 2$ (at
least two genes per genotype) and let $g$ be a fixed gene out of a genotype's
$L/B$ genes. The multiple producers that we consider are the $2^B$ genotypes
that differ from one another only at gene $g$. It follows that every genotype
$i\notin P$ must agree with exactly one genotype $k_i\in P$ at gene $g$ and
differ from all the $2^B-1$ other genotypes in $P$ at this same gene. Therefore,
for some $J_{ik_i}\in\{1,\ldots,L/B-1\}$, genotype $i$ differs from producer
$k_i$ at $J_{ik_i}$ genes and from all other producers at $J_{ik_i}+1$ genes.

Similarly to how we proceeded in Sec.~\ref{sec:sc1}, here we assume either
$\sigma_{ji}=[j\in P][i\notin P]\sigma$ or
$\sigma_{ji}=[j\in P][i\in P][i\neq j]\sigma_\mathrm{p}$, with
$\sigma,\sigma_\mathrm{p}\in[-1,1]$, depending on whether $i\in P$. That is, for
$k\in P$, genotype $i\neq k$ is impacted by the public goods that genotype $k$
produces either through $\sigma_{ki}=\sigma$ (if $i\notin P$) or through
$\sigma_{ki}=\sigma_\mathrm{p}$ (if $i\in P$). We also assume
$\sigma_{kk}=\sigma_\mathrm{pp}$ with $\sigma_\mathrm{pp}\in[-1,1]$ for all
$k\in P$. All $\sigma_{ji}$'s with $j\notin P$ are irrelevant. It follows that
every genotype $k\in P$ has the same value for $x_k$, $d_k$, and $f_k$, which we
henceforth denote by $x_\mathrm{p}$, $d_\mathrm{p}$, and $f_\mathrm{p}$. With
these settings in place, by Eq.~(\ref{di}) we have
\begin{equation}
d_\mathrm{p}=
\sigma_\mathrm{pp}x_\mathrm{p}+(2^B-1)\sigma_\mathrm{p}ux_\mathrm{p}
\end{equation}
and
\begin{equation}
d_i=\sigma u^{J_{k_ii}}x_\mathrm{p}+(2^B-1)\sigma u^{J_{k_ii}+1}x_\mathrm{p}
\end{equation}
for $i\notin P$.

For $\sigma>0$, the fittest genotypes $i\notin P$ are such that
$J_{J_{k_ii}}=1$. Comparing $f_\mathrm{p}$ to $f_+$ depends on how
$\sigma_\mathrm{pp}$ relates to
\begin{equation}
\alpha=
\left[1-(2^B-1)\left(\frac{\sigma_\mathrm{p}}{\sigma}-u\right)\right]\sigma u,
\end{equation}
which for $\sigma_\mathrm{pp}>0$ and $\sigma_\mathrm{p}<[(2^B-1)^{-1}+u]\sigma$
(to ensure $\alpha>0$) extends case~C1 of Sec.~\ref{sec:sc1}. The resulting case
is as follows.
\begin{itemize}
\item[D1.]
$\sigma_\mathrm{pp}>0$, $\sigma>0$, and
$\sigma_\mathrm{p}<[(2^B-1)^{-1}+u]\sigma$:
$f_\mathrm{p}>1$ and $f_+>1$, with three sub-cases:
\begin{itemize}
\item[D1.a.]
$\sigma_\mathrm{pp}>\alpha$:
$f_\mathrm{p}>f_+$;
genotypes $k\in P$ survive.
\item[D1.b.]
$\sigma_\mathrm{pp}=\alpha$:
$f_\mathrm{p}=f_+$;
genotypes $k\in P$ survive, and so do all genotypes $i\notin P$ such that
$J_{k_ii}=1$.
\item[D1.c.]
$\sigma_\mathrm{pp}<\alpha$:
$f_\mathrm{p}<f_+$;
genotypes $i\notin P$ such that $J_{k_ii}=1$ survive.
\end{itemize}
\end{itemize}

Case~C9 can be extended similarly by first noting that for $\sigma<0$ the
fittest genotypes $i\notin P$ are such that $J_{J_{k_ii}}=L/B-1$. It follows
that comparing $f_\mathrm{p}$ to $f_+$ depends on the relationship between
$\sigma_\mathrm{pp}$
and
\begin{equation}
\beta=
\left[1-(2^B-1)\left(\frac{\sigma_\mathrm{p}}{\sigma u^{L/B-2}}-
u\right)\right]\sigma u^{L/B-1}.
\end{equation}
The desired extension comes from assuming $\sigma_\mathrm{pp}<0$ and
$\sigma_\mathrm{p}>[(2^B-1)^{-1}+u]\sigma u^{L/B-2}$ (to ensure $\beta<0$). It
is as follows.
\begin{itemize}
\item[D9.]
$\sigma_\mathrm{pp}<0$, $\sigma<0$, and
$\sigma_\mathrm{p}>[(2^B-1)^{-1}+u]\sigma u^{L/B-2}$:
$f_\mathrm{p}<1$ and $f_+<1$, with three sub-cases:
\begin{itemize}
\item[D9.a.]
$\sigma_\mathrm{pp}>\beta$:
$f_\mathrm{p}>f_+$;
genotypes $k\in P$ survive.
\item[D9.b.]
$\sigma_\mathrm{pp}=\beta$:
$f_\mathrm{p}=f_+$;
genotypes $k\in P$ survive, and so do all genotypes $i\notin P$ such that
$J_{k_ii}=L/B-1$.
\item[D9.c.]
$\sigma_\mathrm{pp}<\beta$:
$f_\mathrm{p}<f_+$;
genotypes $i\notin P$ such that $J_{k_ii}=L/B-1$ survive.
\end{itemize}
\end{itemize}

The trivial cases of Sec.~\ref{sec:sc1} (cases~C2--C8) can be extended
similarly. Of course, allowing for $\sigma_\mathrm{p}$ and $\sigma$ to relate to
each other differently will lead to further variations.

\bibliography{qsensing}

\begin{thebibliography}{43}%
\makeatletter
\providecommand \@ifxundefined [1]{%
 \@ifx{#1\undefined}
}%
\providecommand \@ifnum [1]{%
 \ifnum #1\expandafter \@firstoftwo
 \else \expandafter \@secondoftwo
 \fi
}%
\providecommand \@ifx [1]{%
 \ifx #1\expandafter \@firstoftwo
 \else \expandafter \@secondoftwo
 \fi
}%
\providecommand \natexlab [1]{#1}%
\providecommand \enquote  [1]{``#1''}%
\providecommand \bibnamefont  [1]{#1}%
\providecommand \bibfnamefont [1]{#1}%
\providecommand \citenamefont [1]{#1}%
\providecommand \href@noop [0]{\@secondoftwo}%
\providecommand \href [0]{\begingroup \@sanitize@url \@href}%
\providecommand \@href[1]{\@@startlink{#1}\@@href}%
\providecommand \@@href[1]{\endgroup#1\@@endlink}%
\providecommand \@sanitize@url [0]{\catcode `\\12\catcode `\$12\catcode
  `\&12\catcode `\#12\catcode `\^12\catcode `\_12\catcode `\%12\relax}%
\providecommand \@@startlink[1]{}%
\providecommand \@@endlink[0]{}%
\providecommand \url  [0]{\begingroup\@sanitize@url \@url }%
\providecommand \@url [1]{\endgroup\@href {#1}{\urlprefix }}%
\providecommand \urlprefix  [0]{URL }%
\providecommand \Eprint [0]{\href }%
\providecommand \doibase [0]{https://doi.org/}%
\providecommand \selectlanguage [0]{\@gobble}%
\providecommand \bibinfo  [0]{\@secondoftwo}%
\providecommand \bibfield  [0]{\@secondoftwo}%
\providecommand \translation [1]{[#1]}%
\providecommand \BibitemOpen [0]{}%
\providecommand \bibitemStop [0]{}%
\providecommand \bibitemNoStop [0]{.\EOS\space}%
\providecommand \EOS [0]{\spacefactor3000\relax}%
\providecommand \BibitemShut  [1]{\csname bibitem#1\endcsname}%
\let\auto@bib@innerbib\@empty
\bibitem [{\citenamefont {Miller}\ and\ \citenamefont {Bassler}(2001)}]{mb01}%
  \BibitemOpen
  \bibfield  {author} {\bibinfo {author} {\bibfnamefont {M.~B.}\ \bibnamefont
  {Miller}}\ and\ \bibinfo {author} {\bibfnamefont {B.~L.}\ \bibnamefont
  {Bassler}},\ }\href@noop {} {\bibfield  {journal} {\bibinfo  {journal} {Annu.
  Rev. Microbiol.}\ }\textbf {\bibinfo {volume} {55}},\ \bibinfo {pages} {165}
  (\bibinfo {year} {2001})}\BibitemShut {NoStop}%
\bibitem [{\citenamefont {Hawver}\ \emph {et~al.}(2016)\citenamefont {Hawver},
  \citenamefont {Jung},\ and\ \citenamefont {Ng}}]{hjn16}%
  \BibitemOpen
  \bibfield  {author} {\bibinfo {author} {\bibfnamefont {L.~A.}\ \bibnamefont
  {Hawver}}, \bibinfo {author} {\bibfnamefont {S.~A.}\ \bibnamefont {Jung}},\
  and\ \bibinfo {author} {\bibfnamefont {W.-L.}\ \bibnamefont {Ng}},\
  }\href@noop {} {\bibfield  {journal} {\bibinfo  {journal} {FEMS Microbiol.
  Rev.}\ }\textbf {\bibinfo {volume} {40}},\ \bibinfo {pages} {738} (\bibinfo
  {year} {2016})}\BibitemShut {NoStop}%
\bibitem [{\citenamefont {Papenfort}\ and\ \citenamefont
  {Bassler}(2016)}]{pb16}%
  \BibitemOpen
  \bibfield  {author} {\bibinfo {author} {\bibfnamefont {K.}~\bibnamefont
  {Papenfort}}\ and\ \bibinfo {author} {\bibfnamefont {B.~L.}\ \bibnamefont
  {Bassler}},\ }\href@noop {} {\bibfield  {journal} {\bibinfo  {journal} {Nat.
  Rev. Microbiol.}\ }\textbf {\bibinfo {volume} {14}},\ \bibinfo {pages} {576}
  (\bibinfo {year} {2016})}\BibitemShut {NoStop}%
\bibitem [{\citenamefont {P\'{e}rez-Vel\'{a}zquez}\ \emph
  {et~al.}(2016)\citenamefont {P\'{e}rez-Vel\'{a}zquez}, \citenamefont
  {G{\"{o}}lgeli},\ and\ \citenamefont {Garc\'{\i}a-Contreras}}]{pgg16}%
  \BibitemOpen
  \bibfield  {author} {\bibinfo {author} {\bibfnamefont {J.}~\bibnamefont
  {P\'{e}rez-Vel\'{a}zquez}}, \bibinfo {author} {\bibfnamefont
  {M.}~\bibnamefont {G{\"{o}}lgeli}},\ and\ \bibinfo {author} {\bibfnamefont
  {R.}~\bibnamefont {Garc\'{\i}a-Contreras}},\ }\href@noop {} {\bibfield
  {journal} {\bibinfo  {journal} {Bull. Math. Biol.}\ }\textbf {\bibinfo
  {volume} {78}},\ \bibinfo {pages} {1585} (\bibinfo {year}
  {2016})}\BibitemShut {NoStop}%
\bibitem [{\citenamefont {{da Silva}}\ \emph {et~al.}(2017)\citenamefont {{da
  Silva}}, \citenamefont {Schofield}, \citenamefont {Parsek},\ and\
  \citenamefont {Tseng}}]{sspt17}%
  \BibitemOpen
  \bibfield  {author} {\bibinfo {author} {\bibfnamefont {D.~P.}\ \bibnamefont
  {{da Silva}}}, \bibinfo {author} {\bibfnamefont {M.~C.}\ \bibnamefont
  {Schofield}}, \bibinfo {author} {\bibfnamefont {M.~R.}\ \bibnamefont
  {Parsek}},\ and\ \bibinfo {author} {\bibfnamefont {B.~S.}\ \bibnamefont
  {Tseng}},\ }\href@noop {} {\bibfield  {journal} {\bibinfo  {journal}
  {Pathogens}\ }\textbf {\bibinfo {volume} {6}},\ \bibinfo {pages} {51}
  (\bibinfo {year} {2017})}\BibitemShut {NoStop}%
\bibitem [{\citenamefont {Whiteley}\ \emph {et~al.}(2017)\citenamefont
  {Whiteley}, \citenamefont {Diggle},\ and\ \citenamefont {Greenberg}}]{wdg17}%
  \BibitemOpen
  \bibfield  {author} {\bibinfo {author} {\bibfnamefont {M.}~\bibnamefont
  {Whiteley}}, \bibinfo {author} {\bibfnamefont {S.~P.}\ \bibnamefont
  {Diggle}},\ and\ \bibinfo {author} {\bibfnamefont {E.~P.}\ \bibnamefont
  {Greenberg}},\ }\href@noop {} {\bibfield  {journal} {\bibinfo  {journal}
  {Nature}\ }\textbf {\bibinfo {volume} {551}},\ \bibinfo {pages} {313}
  (\bibinfo {year} {2017})}\BibitemShut {NoStop}%
\bibitem [{\citenamefont {Nealson}\ \emph {et~al.}(1970)\citenamefont
  {Nealson}, \citenamefont {Platt},\ and\ \citenamefont {{Woodland
  Hastings}}}]{nph70}%
  \BibitemOpen
  \bibfield  {author} {\bibinfo {author} {\bibfnamefont {K.~H.}\ \bibnamefont
  {Nealson}}, \bibinfo {author} {\bibfnamefont {T.}~\bibnamefont {Platt}},\
  and\ \bibinfo {author} {\bibfnamefont {J.}~\bibnamefont {{Woodland
  Hastings}}},\ }\href@noop {} {\bibfield  {journal} {\bibinfo  {journal} {J.
  Bacteriol.}\ }\textbf {\bibinfo {volume} {104}},\ \bibinfo {pages} {313}
  (\bibinfo {year} {1970})}\BibitemShut {NoStop}%
\bibitem [{\citenamefont {Greenberg}\ \emph {et~al.}(1979)\citenamefont
  {Greenberg}, \citenamefont {{Woodland Hastings}},\ and\ \citenamefont
  {Ulitzur}}]{ghu79}%
  \BibitemOpen
  \bibfield  {author} {\bibinfo {author} {\bibfnamefont {E.~P.}\ \bibnamefont
  {Greenberg}}, \bibinfo {author} {\bibfnamefont {J.}~\bibnamefont {{Woodland
  Hastings}}},\ and\ \bibinfo {author} {\bibfnamefont {S.}~\bibnamefont
  {Ulitzur}},\ }\href@noop {} {\bibfield  {journal} {\bibinfo  {journal} {Arch.
  Microbiol.}\ }\textbf {\bibinfo {volume} {120}},\ \bibinfo {pages} {87}
  (\bibinfo {year} {1979})}\BibitemShut {NoStop}%
\bibitem [{\citenamefont {Garc\'{\i}a-Bayona}\ and\ \citenamefont
  {Comstock}(2018)}]{gc18}%
  \BibitemOpen
  \bibfield  {author} {\bibinfo {author} {\bibfnamefont {L.}~\bibnamefont
  {Garc\'{\i}a-Bayona}}\ and\ \bibinfo {author} {\bibfnamefont {L.~E.}\
  \bibnamefont {Comstock}},\ }\href@noop {} {\bibfield  {journal} {\bibinfo
  {journal} {Science}\ }\textbf {\bibinfo {volume} {361}},\ \bibinfo {pages}
  {eaat2456} (\bibinfo {year} {2018})}\BibitemShut {NoStop}%
\bibitem [{\citenamefont {Heddi}\ and\ \citenamefont
  {Zaidman-R\'{e}my}(2018)}]{hz18}%
  \BibitemOpen
  \bibfield  {author} {\bibinfo {author} {\bibfnamefont {A.}~\bibnamefont
  {Heddi}}\ and\ \bibinfo {author} {\bibfnamefont {A.}~\bibnamefont
  {Zaidman-R\'{e}my}},\ }\href@noop {} {\bibfield  {journal} {\bibinfo
  {journal} {C. R. Biologies}\ }\textbf {\bibinfo {volume} {341}},\ \bibinfo
  {pages} {290} (\bibinfo {year} {2018})}\BibitemShut {NoStop}%
\bibitem [{\citenamefont {Rutherford}\ and\ \citenamefont
  {Bassler}(2012)}]{rb12}%
  \BibitemOpen
  \bibfield  {author} {\bibinfo {author} {\bibfnamefont {S.~T.}\ \bibnamefont
  {Rutherford}}\ and\ \bibinfo {author} {\bibfnamefont {B.~L.}\ \bibnamefont
  {Bassler}},\ }\href@noop {} {\bibfield  {journal} {\bibinfo  {journal} {Cold
  Spring Harb. Perspect. Med.}\ }\textbf {\bibinfo {volume} {2}},\ \bibinfo
  {pages} {a012427} (\bibinfo {year} {2012})}\BibitemShut {NoStop}%
\bibitem [{\citenamefont {Defoirdt}(2018)}]{d18}%
  \BibitemOpen
  \bibfield  {author} {\bibinfo {author} {\bibfnamefont {T.}~\bibnamefont
  {Defoirdt}},\ }\href@noop {} {\bibfield  {journal} {\bibinfo  {journal}
  {Trends Microbiol.}\ }\textbf {\bibinfo {volume} {26}},\ \bibinfo {pages}
  {313} (\bibinfo {year} {2018})}\BibitemShut {NoStop}%
\bibitem [{\citenamefont {Cornforth}\ \emph {et~al.}(2014)\citenamefont
  {Cornforth}, \citenamefont {Popat}, \citenamefont {McNally}, \citenamefont
  {Gurney}, \citenamefont {Scott-Phillips}, \citenamefont {Ivens},
  \citenamefont {Diggle},\ and\ \citenamefont {Brown}}]{cpmgsidb14}%
  \BibitemOpen
  \bibfield  {author} {\bibinfo {author} {\bibfnamefont {D.~M.}\ \bibnamefont
  {Cornforth}}, \bibinfo {author} {\bibfnamefont {R.}~\bibnamefont {Popat}},
  \bibinfo {author} {\bibfnamefont {L.}~\bibnamefont {McNally}}, \bibinfo
  {author} {\bibfnamefont {J.}~\bibnamefont {Gurney}}, \bibinfo {author}
  {\bibfnamefont {T.~C.}\ \bibnamefont {Scott-Phillips}}, \bibinfo {author}
  {\bibfnamefont {A.}~\bibnamefont {Ivens}}, \bibinfo {author} {\bibfnamefont
  {S.~P.}\ \bibnamefont {Diggle}},\ and\ \bibinfo {author} {\bibfnamefont
  {S.~P.}\ \bibnamefont {Brown}},\ }\href@noop {} {\bibfield  {journal}
  {\bibinfo  {journal} {P. Natl. Acad. Sci. USA}\ }\textbf {\bibinfo {volume}
  {111}},\ \bibinfo {pages} {4280} (\bibinfo {year} {2014})}\BibitemShut
  {NoStop}%
\bibitem [{\citenamefont {Yusufaly}\ and\ \citenamefont
  {Boedicker}(2016)}]{yb16}%
  \BibitemOpen
  \bibfield  {author} {\bibinfo {author} {\bibfnamefont {T.~I.}\ \bibnamefont
  {Yusufaly}}\ and\ \bibinfo {author} {\bibfnamefont {J.~Q.}\ \bibnamefont
  {Boedicker}},\ }\href@noop {} {\bibfield  {journal} {\bibinfo  {journal}
  {Phys. Rev. E}\ }\textbf {\bibinfo {volume} {94}},\ \bibinfo {pages} {062410}
  (\bibinfo {year} {2016})}\BibitemShut {NoStop}%
\bibitem [{\citenamefont {Velasco}\ \emph {et~al.}(2018)\citenamefont
  {Velasco}, \citenamefont {Abkenar}, \citenamefont {Gompper},\ and\
  \citenamefont {Auth}}]{vaga18}%
  \BibitemOpen
  \bibfield  {author} {\bibinfo {author} {\bibfnamefont {C.~A.}\ \bibnamefont
  {Velasco}}, \bibinfo {author} {\bibfnamefont {M.}~\bibnamefont {Abkenar}},
  \bibinfo {author} {\bibfnamefont {G.}~\bibnamefont {Gompper}},\ and\ \bibinfo
  {author} {\bibfnamefont {T.}~\bibnamefont {Auth}},\ }\href@noop {} {\bibfield
   {journal} {\bibinfo  {journal} {Phys. Rev. E}\ }\textbf {\bibinfo {volume}
  {98}},\ \bibinfo {pages} {022605} (\bibinfo {year} {2018})}\BibitemShut
  {NoStop}%
\bibitem [{\citenamefont {Rein}\ \emph {et~al.}(2016)\citenamefont {Rein},
  \citenamefont {Hein\ss}, \citenamefont {Schmid},\ and\ \citenamefont
  {Speck}}]{rhss16}%
  \BibitemOpen
  \bibfield  {author} {\bibinfo {author} {\bibfnamefont {M.}~\bibnamefont
  {Rein}}, \bibinfo {author} {\bibfnamefont {N.}~\bibnamefont {Hein\ss}},
  \bibinfo {author} {\bibfnamefont {F.}~\bibnamefont {Schmid}},\ and\ \bibinfo
  {author} {\bibfnamefont {T.}~\bibnamefont {Speck}},\ }\href@noop {}
  {\bibfield  {journal} {\bibinfo  {journal} {Phys. Rev. Lett.}\ }\textbf
  {\bibinfo {volume} {116}},\ \bibinfo {pages} {058102} (\bibinfo {year}
  {2016})}\BibitemShut {NoStop}%
\bibitem [{\citenamefont {Bauer}\ and\ \citenamefont {Frey}(2018)}]{bf18}%
  \BibitemOpen
  \bibfield  {author} {\bibinfo {author} {\bibfnamefont {M.}~\bibnamefont
  {Bauer}}\ and\ \bibinfo {author} {\bibfnamefont {E.}~\bibnamefont {Frey}},\
  }\href@noop {} {\bibfield  {journal} {\bibinfo  {journal} {Phys. Rev. E}\
  }\textbf {\bibinfo {volume} {97}},\ \bibinfo {pages} {042307} (\bibinfo
  {year} {2018})}\BibitemShut {NoStop}%
\bibitem [{\citenamefont {Russo}\ and\ \citenamefont {Slotine}(2010)}]{rs10}%
  \BibitemOpen
  \bibfield  {author} {\bibinfo {author} {\bibfnamefont {G.}~\bibnamefont
  {Russo}}\ and\ \bibinfo {author} {\bibfnamefont {J.~J.~E.}\ \bibnamefont
  {Slotine}},\ }\href@noop {} {\bibfield  {journal} {\bibinfo  {journal} {Phys.
  Rev. E}\ }\textbf {\bibinfo {volume} {82}},\ \bibinfo {pages} {041919}
  (\bibinfo {year} {2010})}\BibitemShut {NoStop}%
\bibitem [{\citenamefont {Verma}\ \emph {et~al.}(2019)\citenamefont {Verma},
  \citenamefont {Chaurasia},\ and\ \citenamefont {Sinha}}]{vcs19}%
  \BibitemOpen
  \bibfield  {author} {\bibinfo {author} {\bibfnamefont {U.~K.}\ \bibnamefont
  {Verma}}, \bibinfo {author} {\bibfnamefont {S.~S.}\ \bibnamefont
  {Chaurasia}},\ and\ \bibinfo {author} {\bibfnamefont {S.}~\bibnamefont
  {Sinha}},\ }\href@noop {} {\bibfield  {journal} {\bibinfo  {journal} {Phys.
  Rev. E}\ }\textbf {\bibinfo {volume} {100}},\ \bibinfo {pages} {032203}
  (\bibinfo {year} {2019})}\BibitemShut {NoStop}%
\bibitem [{\citenamefont {Fischer}\ \emph {et~al.}(2020)\citenamefont
  {Fischer}, \citenamefont {Schmid},\ and\ \citenamefont {Speck}}]{fss20}%
  \BibitemOpen
  \bibfield  {author} {\bibinfo {author} {\bibfnamefont {A.}~\bibnamefont
  {Fischer}}, \bibinfo {author} {\bibfnamefont {F.}~\bibnamefont {Schmid}},\
  and\ \bibinfo {author} {\bibfnamefont {T.}~\bibnamefont {Speck}},\
  }\href@noop {} {\bibfield  {journal} {\bibinfo  {journal} {Phys. Rev. E}\
  }\textbf {\bibinfo {volume} {101}},\ \bibinfo {pages} {012601} (\bibinfo
  {year} {2020})}\BibitemShut {NoStop}%
\bibitem [{\citenamefont {Singh}\ and\ \citenamefont {{Anil
  Kumar}}(2020)}]{sk20}%
  \BibitemOpen
  \bibfield  {author} {\bibinfo {author} {\bibfnamefont {J.}~\bibnamefont
  {Singh}}\ and\ \bibinfo {author} {\bibfnamefont {A.~V.}\ \bibnamefont {{Anil
  Kumar}}},\ }\href@noop {} {\bibfield  {journal} {\bibinfo  {journal} {Phys.
  Rev. E}\ }\textbf {\bibinfo {volume} {101}},\ \bibinfo {pages} {022606}
  (\bibinfo {year} {2020})}\BibitemShut {NoStop}%
\bibitem [{\citenamefont {Rossell\'{o}-M\'{o}ra}\ and\ \citenamefont
  {Amann}(2015)}]{ra15}%
  \BibitemOpen
  \bibfield  {author} {\bibinfo {author} {\bibfnamefont {R.}~\bibnamefont
  {Rossell\'{o}-M\'{o}ra}}\ and\ \bibinfo {author} {\bibfnamefont
  {R.}~\bibnamefont {Amann}},\ }\href@noop {} {\bibfield  {journal} {\bibinfo
  {journal} {Syst. Appl. Microbiol.}\ }\textbf {\bibinfo {volume} {38}},\
  \bibinfo {pages} {209} (\bibinfo {year} {2015})}\BibitemShut {NoStop}%
\bibitem [{\citenamefont {Angert}(2005)}]{a05}%
  \BibitemOpen
  \bibfield  {author} {\bibinfo {author} {\bibfnamefont {E.~R.}\ \bibnamefont
  {Angert}},\ }\href@noop {} {\bibfield  {journal} {\bibinfo  {journal} {Nat.
  Rev. Microbiol.}\ }\textbf {\bibinfo {volume} {3}},\ \bibinfo {pages} {214}
  (\bibinfo {year} {2005})}\BibitemShut {NoStop}%
\bibitem [{\citenamefont {Bulgherese}(2016)}]{b16}%
  \BibitemOpen
  \bibfield  {author} {\bibinfo {author} {\bibfnamefont {S.}~\bibnamefont
  {Bulgherese}},\ }\href@noop {} {\bibfield  {journal} {\bibinfo  {journal}
  {Environ. Microbiol.}\ }\textbf {\bibinfo {volume} {18}},\ \bibinfo {pages}
  {2305} (\bibinfo {year} {2016})}\BibitemShut {NoStop}%
\bibitem [{\citenamefont {Hall}\ \emph {et~al.}(2017)\citenamefont {Hall},
  \citenamefont {Brockhurst},\ and\ \citenamefont {Harrison}}]{hbh17}%
  \BibitemOpen
  \bibfield  {author} {\bibinfo {author} {\bibfnamefont {J.~P.~J.}\
  \bibnamefont {Hall}}, \bibinfo {author} {\bibfnamefont {M.~A.}\ \bibnamefont
  {Brockhurst}},\ and\ \bibinfo {author} {\bibfnamefont {E.}~\bibnamefont
  {Harrison}},\ }\href@noop {} {\bibfield  {journal} {\bibinfo  {journal}
  {Phil. Trans. R. Soc. B}\ }\textbf {\bibinfo {volume} {372}},\ \bibinfo
  {pages} {20160424} (\bibinfo {year} {2017})}\BibitemShut {NoStop}%
\bibitem [{\citenamefont {Barbosa}\ \emph {et~al.}(2012)\citenamefont
  {Barbosa}, \citenamefont {Donangelo},\ and\ \citenamefont {Souza}}]{bds12}%
  \BibitemOpen
  \bibfield  {author} {\bibinfo {author} {\bibfnamefont {V.~C.}\ \bibnamefont
  {Barbosa}}, \bibinfo {author} {\bibfnamefont {R.}~\bibnamefont {Donangelo}},\
  and\ \bibinfo {author} {\bibfnamefont {S.~R.}\ \bibnamefont {Souza}},\
  }\href@noop {} {\bibfield  {journal} {\bibinfo  {journal} {J. Theor. Biol.}\
  }\textbf {\bibinfo {volume} {312}},\ \bibinfo {pages} {114} (\bibinfo {year}
  {2012})}\BibitemShut {NoStop}%
\bibitem [{\citenamefont {Eigen}(1971)}]{e71}%
  \BibitemOpen
  \bibfield  {author} {\bibinfo {author} {\bibfnamefont {M.}~\bibnamefont
  {Eigen}},\ }\href@noop {} {\bibfield  {journal} {\bibinfo  {journal}
  {Naturwissenschaften}\ }\textbf {\bibinfo {volume} {58}},\ \bibinfo {pages}
  {465} (\bibinfo {year} {1971})}\BibitemShut {NoStop}%
\bibitem [{\citenamefont {Eigen}\ and\ \citenamefont {Schuster}(1977)}]{es77}%
  \BibitemOpen
  \bibfield  {author} {\bibinfo {author} {\bibfnamefont {M.}~\bibnamefont
  {Eigen}}\ and\ \bibinfo {author} {\bibfnamefont {P.}~\bibnamefont
  {Schuster}},\ }\href@noop {} {\bibfield  {journal} {\bibinfo  {journal}
  {Naturwissenschaften}\ }\textbf {\bibinfo {volume} {64}},\ \bibinfo {pages}
  {541} (\bibinfo {year} {1977})}\BibitemShut {NoStop}%
\bibitem [{\citenamefont {Biebricher}\ and\ \citenamefont
  {Eigen}(2006)}]{be06}%
  \BibitemOpen
  \bibfield  {author} {\bibinfo {author} {\bibfnamefont {C.~K.}\ \bibnamefont
  {Biebricher}}\ and\ \bibinfo {author} {\bibfnamefont {M.}~\bibnamefont
  {Eigen}},\ }in\ \href@noop {} {\emph {\bibinfo {booktitle} {Quasispecies:
  Concept and Implications for Virology}}},\ \bibinfo {series} {Current Topics
  in Microbiology and Immunology}, Vol.\ \bibinfo {volume} {299},\ \bibinfo
  {editor} {edited by\ \bibinfo {editor} {\bibfnamefont {E.}~\bibnamefont
  {Domingo}}}\ (\bibinfo  {publisher} {Springer},\ \bibinfo {address} {Berlin,
  Germany},\ \bibinfo {year} {2006})\ pp.\ \bibinfo {pages} {1--31}\BibitemShut
  {NoStop}%
\bibitem [{\citenamefont {Domingo}(2009)}]{d09}%
  \BibitemOpen
  \bibfield  {author} {\bibinfo {author} {\bibfnamefont {E.}~\bibnamefont
  {Domingo}},\ }\href@noop {} {\bibfield  {journal} {\bibinfo  {journal}
  {Contrib. Sci.}\ }\textbf {\bibinfo {volume} {5}},\ \bibinfo {pages} {161}
  (\bibinfo {year} {2009})}\BibitemShut {NoStop}%
\bibitem [{\citenamefont {Lauring}\ and\ \citenamefont {Andino}(2010)}]{la10}%
  \BibitemOpen
  \bibfield  {author} {\bibinfo {author} {\bibfnamefont {A.~S.}\ \bibnamefont
  {Lauring}}\ and\ \bibinfo {author} {\bibfnamefont {R.}~\bibnamefont
  {Andino}},\ }\href@noop {} {\bibfield  {journal} {\bibinfo  {journal} {PLoS
  Pathog.}\ }\textbf {\bibinfo {volume} {6}},\ \bibinfo {pages} {e1001005}
  (\bibinfo {year} {2010})}\BibitemShut {NoStop}%
\bibitem [{\citenamefont {M\'{a}s}\ \emph {et~al.}(2010)\citenamefont
  {M\'{a}s}, \citenamefont {L\'{o}pez-Gal\'{\i}ndez}, \citenamefont {Cacho},
  \citenamefont {G\'{o}mez},\ and\ \citenamefont {Mart\'{\i}nez}}]{mlcgm10}%
  \BibitemOpen
  \bibfield  {author} {\bibinfo {author} {\bibfnamefont {A.}~\bibnamefont
  {M\'{a}s}}, \bibinfo {author} {\bibfnamefont {C.}~\bibnamefont
  {L\'{o}pez-Gal\'{\i}ndez}}, \bibinfo {author} {\bibfnamefont
  {I.}~\bibnamefont {Cacho}}, \bibinfo {author} {\bibfnamefont
  {J.}~\bibnamefont {G\'{o}mez}},\ and\ \bibinfo {author} {\bibfnamefont
  {M.~A.}\ \bibnamefont {Mart\'{\i}nez}},\ }\href@noop {} {\bibfield  {journal}
  {\bibinfo  {journal} {J. Mol. Biol.}\ }\textbf {\bibinfo {volume} {397}},\
  \bibinfo {pages} {865} (\bibinfo {year} {2010})}\BibitemShut {NoStop}%
\bibitem [{\citenamefont {Barbosa}\ \emph {et~al.}(2015)\citenamefont
  {Barbosa}, \citenamefont {Donangelo},\ and\ \citenamefont {Souza}}]{bds15}%
  \BibitemOpen
  \bibfield  {author} {\bibinfo {author} {\bibfnamefont {V.~C.}\ \bibnamefont
  {Barbosa}}, \bibinfo {author} {\bibfnamefont {R.}~\bibnamefont {Donangelo}},\
  and\ \bibinfo {author} {\bibfnamefont {S.~R.}\ \bibnamefont {Souza}},\
  }\href@noop {} {\bibfield  {journal} {\bibinfo  {journal} {J. Stat. Mech.}\
  ,\ \bibinfo {pages} {P01022}} (\bibinfo {year} {2015})}\BibitemShut {NoStop}%
\bibitem [{\citenamefont {Barbosa}\ \emph {et~al.}(2016)\citenamefont
  {Barbosa}, \citenamefont {Donangelo},\ and\ \citenamefont {Souza}}]{bds16}%
  \BibitemOpen
  \bibfield  {author} {\bibinfo {author} {\bibfnamefont {V.~C.}\ \bibnamefont
  {Barbosa}}, \bibinfo {author} {\bibfnamefont {R.}~\bibnamefont {Donangelo}},\
  and\ \bibinfo {author} {\bibfnamefont {S.~R.}\ \bibnamefont {Souza}},\
  }\href@noop {} {\bibfield  {journal} {\bibinfo  {journal} {J. Stat. Mech.}\
  ,\ \bibinfo {pages} {063501}} (\bibinfo {year} {2016})}\BibitemShut {NoStop}%
\bibitem [{\citenamefont {Barbosa}\ \emph {et~al.}(2018)\citenamefont
  {Barbosa}, \citenamefont {Donangelo},\ and\ \citenamefont {Souza}}]{bds18}%
  \BibitemOpen
  \bibfield  {author} {\bibinfo {author} {\bibfnamefont {V.~C.}\ \bibnamefont
  {Barbosa}}, \bibinfo {author} {\bibfnamefont {R.}~\bibnamefont {Donangelo}},\
  and\ \bibinfo {author} {\bibfnamefont {S.~R.}\ \bibnamefont {Souza}},\
  }\href@noop {} {\bibfield  {journal} {\bibinfo  {journal} {Phys. Rev. E}\
  }\textbf {\bibinfo {volume} {98}},\ \bibinfo {pages} {032409} (\bibinfo
  {year} {2018})}\BibitemShut {NoStop}%
\bibitem [{\citenamefont {Sokolovskaya}\ \emph {et~al.}(2020)\citenamefont
  {Sokolovskaya}, \citenamefont {Shelton},\ and\ \citenamefont {Taga}}]{sst20}%
  \BibitemOpen
  \bibfield  {author} {\bibinfo {author} {\bibfnamefont {O.~M.}\ \bibnamefont
  {Sokolovskaya}}, \bibinfo {author} {\bibfnamefont {A.~N.}\ \bibnamefont
  {Shelton}},\ and\ \bibinfo {author} {\bibfnamefont {M.~E.}\ \bibnamefont
  {Taga}},\ }\href@noop {} {\bibfield  {journal} {\bibinfo  {journal}
  {Science}\ }\textbf {\bibinfo {volume} {369}},\ \bibinfo {pages} {eaba0165}
  (\bibinfo {year} {2020})}\BibitemShut {NoStop}%
\bibitem [{\citenamefont {Xavier}\ \emph {et~al.}(2011)\citenamefont {Xavier},
  \citenamefont {Kim},\ and\ \citenamefont {Foster}}]{xkf11}%
  \BibitemOpen
  \bibfield  {author} {\bibinfo {author} {\bibfnamefont {J.~B.}\ \bibnamefont
  {Xavier}}, \bibinfo {author} {\bibfnamefont {W.}~\bibnamefont {Kim}},\ and\
  \bibinfo {author} {\bibfnamefont {K.~R.}\ \bibnamefont {Foster}},\
  }\href@noop {} {\bibfield  {journal} {\bibinfo  {journal} {Mol. Microbiol.}\
  }\textbf {\bibinfo {volume} {79}},\ \bibinfo {pages} {166} (\bibinfo {year}
  {2011})}\BibitemShut {NoStop}%
\bibitem [{\citenamefont {Dandekar}\ \emph {et~al.}(2012)\citenamefont
  {Dandekar}, \citenamefont {Chugani},\ and\ \citenamefont
  {Greenberg}}]{dcg12}%
  \BibitemOpen
  \bibfield  {author} {\bibinfo {author} {\bibfnamefont {A.~A.}\ \bibnamefont
  {Dandekar}}, \bibinfo {author} {\bibfnamefont {S.}~\bibnamefont {Chugani}},\
  and\ \bibinfo {author} {\bibfnamefont {E.~P.}\ \bibnamefont {Greenberg}},\
  }\href@noop {} {\bibfield  {journal} {\bibinfo  {journal} {Science}\ }\textbf
  {\bibinfo {volume} {338}},\ \bibinfo {pages} {264} (\bibinfo {year}
  {2012})}\BibitemShut {NoStop}%
\bibitem [{\citenamefont {Majerczyk}\ \emph {et~al.}(2016)\citenamefont
  {Majerczyk}, \citenamefont {Schneider},\ and\ \citenamefont
  {Greenberg}}]{msg16}%
  \BibitemOpen
  \bibfield  {author} {\bibinfo {author} {\bibfnamefont {C.}~\bibnamefont
  {Majerczyk}}, \bibinfo {author} {\bibfnamefont {E.}~\bibnamefont
  {Schneider}},\ and\ \bibinfo {author} {\bibfnamefont {E.~P.}\ \bibnamefont
  {Greenberg}},\ }\href@noop {} {\bibfield  {journal} {\bibinfo  {journal}
  {eLife}\ }\textbf {\bibinfo {volume} {5}},\ \bibinfo {pages} {e14712}
  (\bibinfo {year} {2016})}\BibitemShut {NoStop}%
\bibitem [{\citenamefont {Hornby}\ \emph {et~al.}(2001)\citenamefont {Hornby},
  \citenamefont {Jensen}, \citenamefont {Lisec}, \citenamefont {Tasto},
  \citenamefont {Jahnke}, \citenamefont {Shoemaker}, \citenamefont {Dussault},\
  and\ \citenamefont {Nickerson}}]{hjltjsdn01}%
  \BibitemOpen
  \bibfield  {author} {\bibinfo {author} {\bibfnamefont {J.~M.}\ \bibnamefont
  {Hornby}}, \bibinfo {author} {\bibfnamefont {E.~C.}\ \bibnamefont {Jensen}},
  \bibinfo {author} {\bibfnamefont {A.~D.}\ \bibnamefont {Lisec}}, \bibinfo
  {author} {\bibfnamefont {J.~J.}\ \bibnamefont {Tasto}}, \bibinfo {author}
  {\bibfnamefont {B.}~\bibnamefont {Jahnke}}, \bibinfo {author} {\bibfnamefont
  {R.}~\bibnamefont {Shoemaker}}, \bibinfo {author} {\bibfnamefont
  {P.}~\bibnamefont {Dussault}},\ and\ \bibinfo {author} {\bibfnamefont
  {K.~W.}\ \bibnamefont {Nickerson}},\ }\href@noop {} {\bibfield  {journal}
  {\bibinfo  {journal} {Appl. Environ. Microbiol.}\ }\textbf {\bibinfo {volume}
  {67}},\ \bibinfo {pages} {2982} (\bibinfo {year} {2001})}\BibitemShut
  {NoStop}%
\bibitem [{\citenamefont {Erez}\ \emph {et~al.}(2017)\citenamefont {Erez},
  \citenamefont {Steinberger-Levy}, \citenamefont {Shamir}, \citenamefont
  {Doron}, \citenamefont {Stokar-Avihail}, \citenamefont {Peleg}, \citenamefont
  {Melamed}, \citenamefont {Leavitt}, \citenamefont {Savidor}, \citenamefont
  {Albeck}, \citenamefont {Amitai},\ and\ \citenamefont
  {Sorek}}]{essdspmlsaas17}%
  \BibitemOpen
  \bibfield  {author} {\bibinfo {author} {\bibfnamefont {Z.}~\bibnamefont
  {Erez}}, \bibinfo {author} {\bibfnamefont {I.}~\bibnamefont
  {Steinberger-Levy}}, \bibinfo {author} {\bibfnamefont {M.}~\bibnamefont
  {Shamir}}, \bibinfo {author} {\bibfnamefont {S.}~\bibnamefont {Doron}},
  \bibinfo {author} {\bibfnamefont {A.}~\bibnamefont {Stokar-Avihail}},
  \bibinfo {author} {\bibfnamefont {Y.}~\bibnamefont {Peleg}}, \bibinfo
  {author} {\bibfnamefont {S.}~\bibnamefont {Melamed}}, \bibinfo {author}
  {\bibfnamefont {A.}~\bibnamefont {Leavitt}}, \bibinfo {author} {\bibfnamefont
  {A.}~\bibnamefont {Savidor}}, \bibinfo {author} {\bibfnamefont
  {S.}~\bibnamefont {Albeck}}, \bibinfo {author} {\bibfnamefont
  {G.}~\bibnamefont {Amitai}},\ and\ \bibinfo {author} {\bibfnamefont
  {R.}~\bibnamefont {Sorek}},\ }\href@noop {} {\bibfield  {journal} {\bibinfo
  {journal} {Nature}\ }\textbf {\bibinfo {volume} {541}},\ \bibinfo {pages}
  {488} (\bibinfo {year} {2017})}\BibitemShut {NoStop}%
\bibitem [{\citenamefont {Buddingh$'$}\ \emph {et~al.}(2020)\citenamefont
  {Buddingh$'$}, \citenamefont {Elzinga},\ and\ \citenamefont {{van
  Hest}}}]{beh20}%
  \BibitemOpen
  \bibfield  {author} {\bibinfo {author} {\bibfnamefont {B.~C.}\ \bibnamefont
  {Buddingh$'$}}, \bibinfo {author} {\bibfnamefont {J.}~\bibnamefont
  {Elzinga}},\ and\ \bibinfo {author} {\bibfnamefont {J.~C.~M.}\ \bibnamefont
  {{van Hest}}},\ }\href@noop {} {\bibfield  {journal} {\bibinfo  {journal}
  {Nat. Commun.}\ }\textbf {\bibinfo {volume} {11}},\ \bibinfo {pages} {1652}
  (\bibinfo {year} {2020})}\BibitemShut {NoStop}%
\bibitem [{\citenamefont {Afek}\ \emph {et~al.}(2011)\citenamefont {Afek},
  \citenamefont {Alon}, \citenamefont {Barad}, \citenamefont {Homstein},
  \citenamefont {Barkai},\ and\ \citenamefont {Bar-Joseph}}]{aabhbb11}%
  \BibitemOpen
  \bibfield  {author} {\bibinfo {author} {\bibfnamefont {Y.}~\bibnamefont
  {Afek}}, \bibinfo {author} {\bibfnamefont {N.}~\bibnamefont {Alon}}, \bibinfo
  {author} {\bibfnamefont {O.}~\bibnamefont {Barad}}, \bibinfo {author}
  {\bibfnamefont {E.}~\bibnamefont {Homstein}}, \bibinfo {author}
  {\bibfnamefont {N.}~\bibnamefont {Barkai}},\ and\ \bibinfo {author}
  {\bibfnamefont {Z.}~\bibnamefont {Bar-Joseph}},\ }\href@noop {} {\bibfield
  {journal} {\bibinfo  {journal} {Science}\ }\textbf {\bibinfo {volume}
  {331}},\ \bibinfo {pages} {183} (\bibinfo {year} {2011})}\BibitemShut
  {NoStop}%
\end{thebibliography}%
\bibliographystyle{apsrev4-2}

\end{document}